\documentclass{llncs}

\usepackage[table]{xcolor}
\usepackage{tikz}

\usepackage{amsmath,amssymb}
\usepackage{rotating}
\usepackage{appendix}
\usepackage{colortbl}
\usepackage[normalem]{ulem}
\usepackage{enumitem}
\usepackage{orcidlink}

\usepackage{textcomp,marvosym}

\usepackage{nameref,hyperref}

\usepackage{xargs}                      

\usepackage{array}

\usepackage{listings} 

\usepackage{subcaption}

\usepackage{algorithmicx}
\usepackage[noend]{algpseudocode}
\usepackage{algorithm}
\usepackage{diagbox}
\algdef{SE}[DOWHILE]{Do}{doWhile}{\algorithmicdo}[1]{\algorithmicwhile\ #1}%

\usepackage{adjustbox}

\usepackage{xr}
\usepackage{catchfilebetweentags}

\usepackage{tkz-orm}
\usetikzlibrary{arrows,shapes,automata,backgrounds,petri,decorations.markings}
\usetikzlibrary{matrix,positioning,decorations.pathreplacing,calc,tikzmark}

\tikzset{
  ln/.style  = { draw, thick, fill=black, circle, inner sep=0.3mm},
  sqloc/.style  = { draw, thick, inner sep=1.0mm },
  loc/.style    = { draw, circle, thick, inner sep=1.0mm },
  notice/.style= { draw, rectangle callout, thick, rounded corners=5pt,fill=blue!10,callout relative pointer={#1} },
  treenode/.style = {align=center, inner sep=0pt, text centered,
    font=\sffamily},
  arn_n/.style = {treenode, rectangle, font=\sffamily\bfseries },
  mymat/.style = { 
    left delimiter={[}, right delimiter ={]},nodes={anchor=base east} },
  align at top/.style={baseline=(current bounding box.north)},
  stack/.style={rectangle split, rectangle split parts=#1,draw, anchor=center}
}

\usepackage{verbatim}
\usepackage{pgfplots}
	
\def\cca#1{\cellcolor{blue!#10}\ifnum #1>5\color{white}\fi{#1}}

\usepackage{tikz}
\usetikzlibrary{tikzmark,decorations.pathreplacing,arrows,shapes,positioning,shadows,trees,shapes.gates.logic.US,arrows.meta,shapes,automata,petri,calc}
\usepackage{caption}
\usepackage{adjustbox}
\usepackage[colorinlistoftodos,prependcaption,textsize=tiny]{todonotes}
\newcommandx{\unsure}[2][1=]{\todo[linecolor=red,backgroundcolor=red!25,bordercolor=red,#1]{#2}}
\newcommandx{\change}[2][1=]{\todo[linecolor=blue,backgroundcolor=blue!25,bordercolor=blue,#1]{#2}}
\newcommandx{\info}[2][1=]{\todo[linecolor=OliveGreen,backgroundcolor=OliveGreen!25,bordercolor=OliveGreen,#1]{#2}}
\newcommandx{\improve}[2][1=]{\todo[linecolor=Plum,backgroundcolor=Plum!25,bordercolor=Plum,#1]{#2}}
\usepackage{verbatim}
\usepackage{scalefnt}
\usepackage{wrapfig}
\usepackage{tikz-qtree}

\mathchardef\mhyphen="2D
\newcommand{\shorteq}{%
  \settowidth{\@tempdima}{-}
  \resizebox{\@tempdima}{\height}{=}%
}

\colorlet{fv}{gray!55}
\colorlet{ai}{gray!10}
\colorlet{ar}{gray!38}

\usepackage[edges]{forest}
\usepackage[T1]{fontenc}

\tikzset{%
  parent/.style={align=center,text width=3cm,rounded corners=3pt},
  child/.style={align=center,text width=3cm,rounded corners=3pt},
}

\def\addlegendimage{\csname pgfplots@addlegendimage\endcsname}

 \usepackage{tikz}
 \usepackage{xparse}
\pgfdeclarelayer{myback}
\pgfsetlayers{myback,background,main}
\usetikzlibrary{tikzmark,decorations.pathreplacing,arrows,shapes,positioning,shadows,trees,shapes.gates.logic.US,arrows.meta,shapes,automata,petri,calc,matrix,backgrounds}
\tikzset{mycolor/.style = {line width=1bp,color=#1}}%
\tikzset{myfillcolor/.style = {draw,fill=#1}}%

\NewDocumentCommand{\highlight}{O{blue!40} m m}{%
\draw[mycolor=#1] (#2.north west)rectangle (#3.south east);
}

\NewDocumentCommand{\fhighlight}{O{blue!40} m m}{%
\draw[myfillcolor=#1] (#2.north west)rectangle (#3.south east);
}
 \usetikzlibrary{matrix,decorations.pathreplacing, calc, positioning}

\newcolumntype{+}{!{\vrule width 2pt}}

\newlength\savedwidth

\usepackage{array}
\newcolumntype{L}[1]{>{\raggedright\let\newline\\\arraybackslash\hspace{0pt}}m{#1}}
\newcolumntype{C}[1]{>{\centering\let\newline\\\arraybackslash\hspace{0pt}}m{#1}}
\newcolumntype{R}[1]{>{\raggedleft\let\newline\\\arraybackslash\hspace{0pt}}m{#1}}

\newcommand{\limplies}{\Rightarrow}

\newcommand{\Land}{\bigwedge}
\newcommand{\Lor}{\bigvee}

\newcommand{\lequiv}{\Leftrightarrow}

\newcommand{\relu}{\textsc{ReLU}}

\newcommand{\alphabeta}{$\alpha\beta$-CROWN}

\newcommand{\marabou}{\textsc{marabou}}
\newcommand{\pyrat}{\textsc{PyRAT}}

\newcommand{\safe}{\textsc{safe}}
\newcommand{\unsafe}{\textsc{unsafe}}
\newcommand{\timeout}{\textsc{timeout}}

\newcommand{\softmax}{\textsc{softmax}}
\newcommand{\Conf}{\textsc{Conf}}
\newcommand{\dist}{\mathtt{dist}}
\newcommand{\gi}{x^*}

\newcommand{\vnnlib}{\textsc{vnnlib}}

\newcommand{\vnncomp}{VNN-COMP}
\newcommand{\smallparam}{\eta}
\newcommand{\thresh}{\tau}

\newtheorem{customclaim}{Claim}

\title{Formal Reasoning About Confidence and Automated Verification of Neural Networks}
\author{Mohammad Afzal\inst{1,2}\orcidlink{0000-0002-6173-3959} 
      \and S. Akshay\inst{1}\orcidlink{0000-0002-2471-5997}
      \and Blaise Genest\inst{3,4}\orcidlink{0000-0002-5758-1876}
      \and Ashutosh Gupta\inst{1}\orcidlink{0009-0003-7755-2006}
      }

\institute{Indian Institute of Technology Bombay, Mumbai, India\and TCS Research, Pune, India\and CNRS@CREATE \& IPAL, Singapore\and CNRS, IPAL, France}

\begin{document}

\maketitle

\begin{abstract}
  In the last decade, a large body of work has emerged on robustness of neural networks, i.e., checking if the decision remains unchanged when the input is slightly perturbed. However, most of these approaches ignore the confidence of a neural network on its output. In this work, we aim to develop a generalized framework for formally reasoning about the confidence along with robustness in neural networks. We propose a simple yet expressive grammar that captures various confidence-based specifications. We develop a novel and unified technique to verify all instances of the grammar in a homogeneous way, viz., by adding a few additional layers to the neural network, which enables the use any state-of-the-art neural network verification tool. We perform an extensive experimental evaluation over a large suite of 8870 benchmarks, where the largest network has 138M parameters, and show that this outperforms ad-hoc encoding approaches by a significant margin.
\end{abstract}

\section{Introduction}
\label{sec:intro}
Neural networks are increasingly being used in safety-critical applications such as autonomous vehicles, medical diagnosis, and speech recognition~\cite{bojarski2016end,amato2013artificial,hinton2012deep}. However, as is well known e.g., in the image classification setting ~\cite{goodfellow2014explaining}, a small perturbation in the input image may lead to misclassifying the output of the network, even if the perturbed image looks exactly like the original image. Such examples are called adversarial examples and there is a long line of work~\cite{dunn2021exposing,hayes2018learning,baluja2018learning} that focuses on generating such examples. But failure to generate adversarial examples does not guarantee that they do not exist, and one often resorts to formal methods techniques for such guarantees.

One such paradigm has emerged around local (i.e., around a given input) robustness verification for image classification, where different techniques have been developed, including sound but incomplete techniques such as~\cite{dvijotham2018dual,singh2018boosting,weng2018towards,zhang2018efficient} as well as less scalable but complete techniques such as~\cite{katz2017reluplex,ehlers2017formal,huang2017safety,wang2021beta,xu2020fast}. The success of these approaches can be judged by the \vnncomp{}~competition for robustness verification of neural networks~\cite{brix2024fifth}, which 
has several competing solvers working on neural networks of fairly large size, with the largest network having approximately $13.16$ Million activations.

\begin{figure}[th!]
  \centering
  \begin{minipage}{0.95\textwidth}
    \centering
    \begin{tabular}{cccc}
      \scriptsize \texttt{horse:85.64}  & \scriptsize \texttt{deer:24.13} & \scriptsize \texttt{airplane:88.87} & \scriptsize \texttt{automobile:22.13} \\
      \scriptsize \texttt{dog : 9.36} & \scriptsize \texttt{horse:20.42} & \scriptsize \texttt{bird:6.13}  & \scriptsize \texttt{airplane:21.42}  \\
      \includegraphics[scale=0.05]{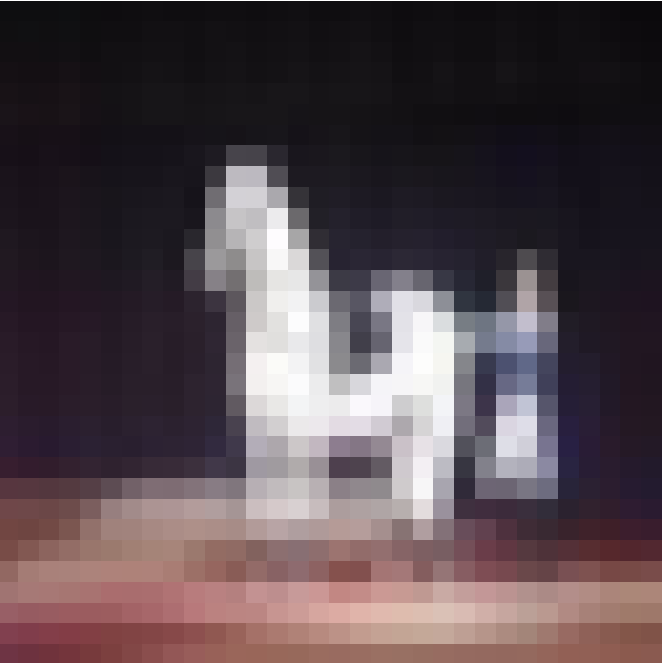} & \includegraphics[scale=0.05]{figs/mot_images/relaxed/im_horse_85.64_deer_24.13_a.png} &
      \includegraphics[scale=0.053]{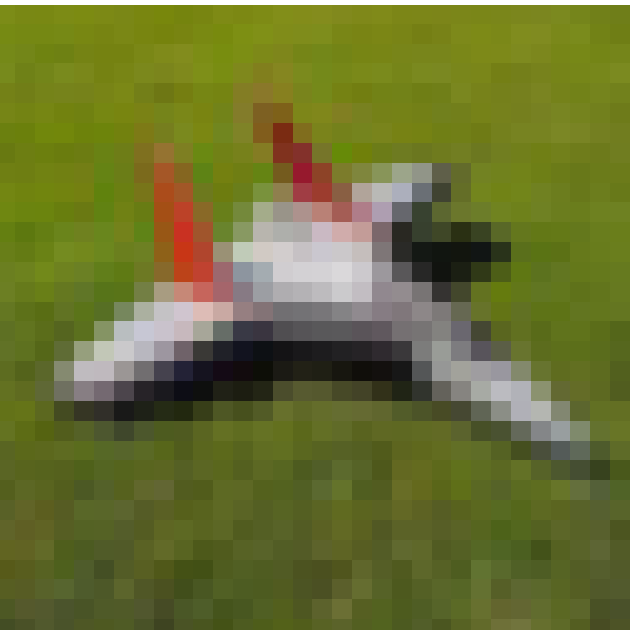} & \includegraphics[scale=0.053]{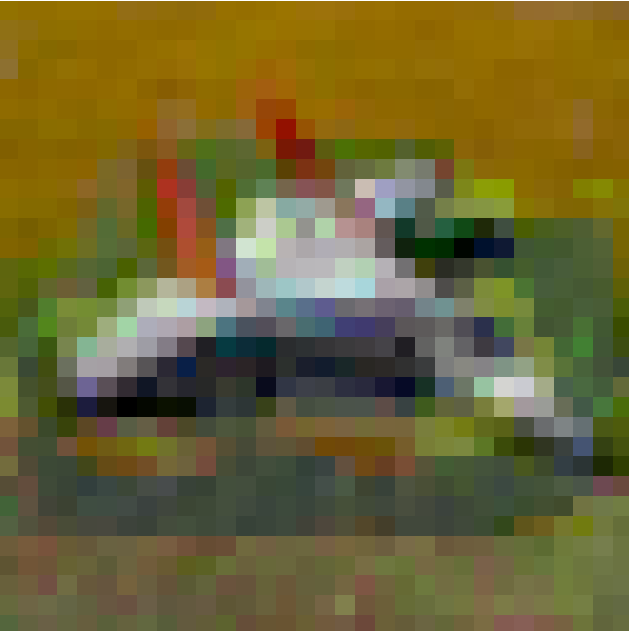}
    \end{tabular}

    \hspace{-.3in} (a) \hspace{1in} (b)
  \end{minipage}
  \begin{minipage}{0.95\textwidth}
    \centering
    \begin{tabular}{cccccc}
      \scriptsize \texttt{ship:96.17} & \scriptsize \texttt{ship:22.21} & \scriptsize \texttt{horse:97.72} & \scriptsize \texttt{horse:27.85} & \scriptsize \begin{tabular}{@{}c@{}}\texttt{label:7}\\ \texttt{label:9} \end{tabular}  & \scriptsize \begin{tabular}{@{}c@{}} \texttt{label:9}\\ \texttt{label:7} \end{tabular}   \\
       \includegraphics[scale=0.05]{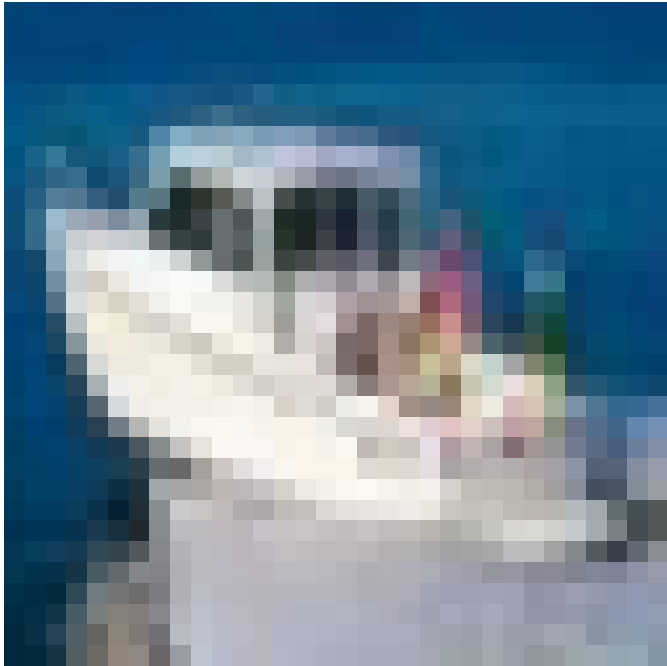} & \includegraphics[scale=0.05]{figs/mot_images/strong/im_ship_96.17_ship_22.21_a.png} &
      \includegraphics[scale=0.05]{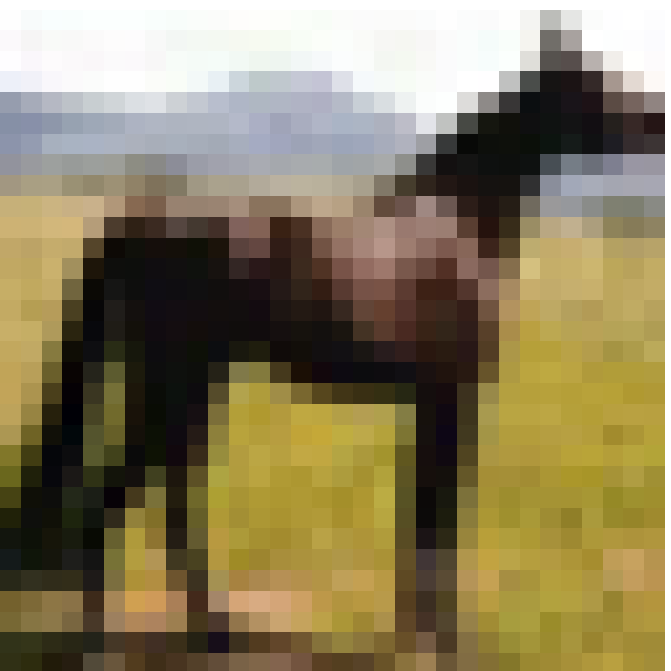} & \includegraphics[scale=0.05]{figs/mot_images/strong/im_horse_97.72_horse_27.85_a.png} & \includegraphics[scale=0.05]{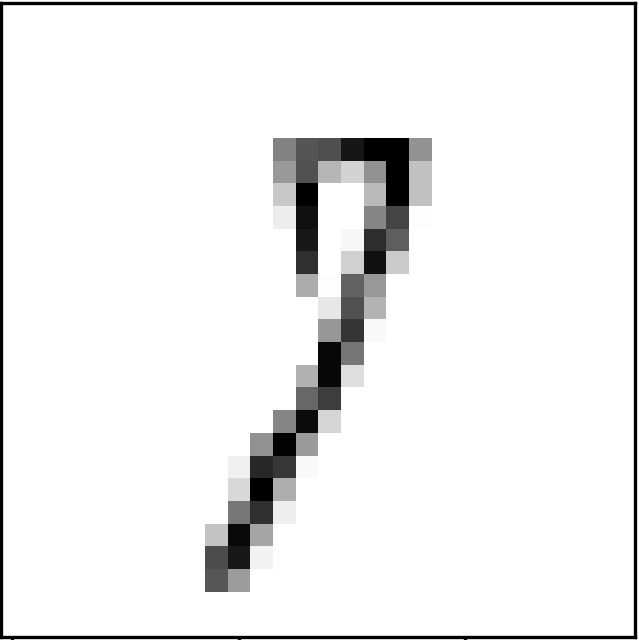} & \includegraphics[scale=0.05]{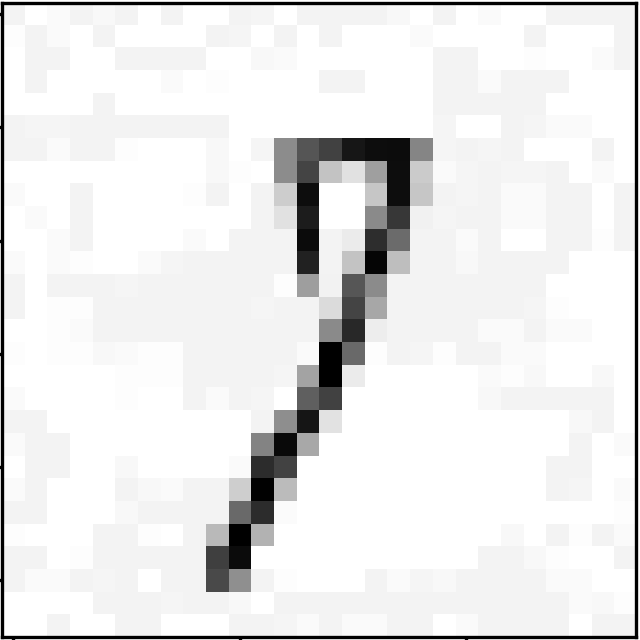} 
    \end{tabular}

    (c) \hspace{.8in} (d) \hspace{.8in} (e)
  \end{minipage}
    
  
  \caption{
    \textbf{Relaxed Robustness: (a-b)} The network \texttt{convBigRELU-PGD.onnx} correctly classified the original image (left) as \texttt{horse} (and resp. \texttt{airplane}) with high confidence. With an input perturbation of $16/255$, we can find misclassified images (right) but with low confidence. In fact, it turns out all counterexamples have low confidence and hence verification succeeds under the relaxed robustness criterion with an $80\%$ confidence threshold, while state-of-the-art verifiers would have declared this network non-robust. \textbf{Strong robustness: (c-d)}  The network \texttt{convBigRELU-PGD.onnx} classified the original image (left) of class \texttt{ship/horse/deer} with very high confidence and we found images (right) within perturbation of $16/255$, such that the confidence drops drastically, although the class remains same. These images are robust with respect to the standard and relaxed robust criteria but not robust with respect to the strong robust criteria if confidence is allowed to fall upto $30\%$. \textbf{Top-$k$ robustness: (e)} The network \texttt{mnist-net-256x6.onnx} correctly classifies the original input as label \(7\). With an input perturbation of \(13/255\), we obtain a counterexample classified as label \(9\). Top-\(k\) robustness requires that the top-\(k\) predictions remain unchanged under input perturbations. For \(k = 2\), the top-2 predictions are labels \(7\) and \(9\), and this set is the same for both the original and perturbed images. Hence, the image shown above satisfies top-2 robustness but not local robustness.
  }
  \label{fig:mot:combine}
  \vspace{-5mm}
\end{figure}

\noindent {\it Different Robustness Variants. }
Different applications require different variants of robustness (e.g., see~\cite{casadio2022neural}). Further, many existing works on robustness verification regard answers returned by classifiers as binary, ignoring the fact that classifiers provide a confidence in the counterexample in terms of the classification probability, which is computed by the well known \softmax{} function~\cite{ia2016deep}. For instance, in Figure~\ref{fig:mot:combine}(a-b) an image from the CIFAR-10~\cite{krizhevsky2009learning} dataset, is misclassified after an input perturbation, but with a very low confidence. If all misclassifications are low confidence ones, should such a network be called non-robust? This motivates a more {\em relaxed} notion of robustness which would allow such examples. At the other extreme, a {\em stronger} notion of robustness says that the network in Figure~\ref{fig:mot:combine}(c-e) should be called non-robust since under minor input perturbations, the confidence in the classification has vastly changed, even if no misclassification is reported. Finally, we may consider top-$k$ robustness~\cite{leino2021relaxing} that requires the top-$k$ predictions to remain within the top-$k$ set under input perturbations, as illustrated in Figure~\ref{fig:mot:combine}(f).
To handle variants such as those described above, the first step is to introduce a common specification mechanism for all of them. In this work, we use a simple grammar to provide a generalized way to reason about the confidence of neural network outputs. In brief, this grammar captures Boolean combinations of confidence-based and non-confidence-based conditions, and the variants introduced earlier are specific instances of this expressive framework. Since our grammar involved confidence, it becomes necessary to approximate the confidence (\softmax{}) function to analyze these properties effectively, which we do, with formal guarantees on the error.

\noindent {\em Difficulties in encoding.} The main difficulty that remains is whether such rich properties modeled as Boolean combinations of linear inequalities can directly be encoded in state-of-the-art tools. As standardized in \vnncomp{}, robustness properties are specified in a special \vnnlib{} format as pre- and post-conditions that a neural network must satisfy. Although \vnnlib{} supports arbitrary Boolean combinations of linear constraints, most state-of-the-art neural network verifiers are optimized for simpler post-conditions which typically involve only disjunctions or conjunctions of linear atoms of over the outputs. Thus, verifying properties with multi-layered conjunctions, disjunctions, or combinations of both poses significant challenges:  
(i) Modifying the verifier's code to handle complex post-conditions requires a deep understanding of its implementation, which can be challenging for users unfamiliar with the codebase, i.e. \alphabeta{}~\cite{alphabetacrown} (ii) The source code for some commercial verifiers may not be publicly available, restricting the ability to make modifications. (iii) Even when access is available, adapting the code for each new property format is time-consuming and prone to errors. (iv) Constraint-based tools, i.e., \marabou{}~\cite{wu2024marabou}, allow such properties to be encoded directly as constraints, but users are then limited to a particular solver only. (v) Advanced techniques like Projected Gradient Descent (PGD) attacks, which are highly effective at finding counterexamples in neural networks, cannot always be applied directly due to the complexity of certain post-conditions.

\noindent{\it Simplifying post-conditions by adding layers.}
To overcome these challenges, we propose a technique that simplifies arbitrary postconditions by appending a few additional layers to the neural network. Intuitively, we encode an arbitrary postcondition as a sequence of layers and attach these layers to the original network. Our method is inspired by the NP-completeness proof in~\cite{katz2021reluplex}, where a reduction from 3-SAT formulas to neural networks is constructed. However, their encoding requires real variables to take values close to either 0 or 1, whereas our framework generalizes this idea to handle arbitrary formulas expressed in Linear Real Arithmetic (LRA) form. Our transformation converts the post-condition into a straightforward form, such as $y \geq 0$ or $y > 0$, where $y$ is the output of an added node in the neural network, while maintaining low error bounds. 
This intricate transformation generates new layers that encode parts of the formulas and composes the outputs of these layers according to the Boolean operations in the formulas. 
Since most verifiers support the \relu{} activation function, we employ \relu{} to model Boolean operations in the post-conditions. The output of a sum of ReLUs can model conjunction or disjunction: if all inputs are negative, the output is zero; if any input is positive, the output is positive. Since we use ReLU operators to model all Boolean operations, conjunctions and disjunctions interpret input signals in opposite ways i.e., for a conjunction, a negative input yields 1 and a positive input yields 0, while for a disjunction, a positive input yields 0 and a negative input yields 1. To enable composition of conjunction outputs into disjunctions and vice versa, we introduce a novel technique that reverses the outputs using a flip operation, while maintaining low error bounds. These simplified post-conditions can be verified using any state-of-the-art verifier as a black box, eliminating the need for code modifications and ensuring seamless integration with existing tools.
Summarizing, {\em our main contributions} are:
\begin{itemize}
  \item We define a grammar that supports generalized reasoning on confidence of the outputs of the neural network. This grammar captures existing notions from the literature such as strong~\cite{casadio2022neural} and top-k robustness~\cite{leino2021relaxing}.
  \item Our framework allows us to define novel notions of relaxed robustness that incorporate confidence by distinguishing low-confidence counterexamples during robustness analysis.
  \item To encode the \softmax{} function (characterizing the confidence) into Boolean combinations of linear constraints, we approximate it with formal guarantees on the error incurred by the approximation.
  \item We provide a new technique to verify any instance of the grammar, 
   via a general encoding (for all properties that can be expressed using this grammar), which  adds layers to the neural network. Our encoding enables the use of state-of-the-art neural network verifiers like \alphabeta{}, \pyrat{}.
  \item We perform a wide set of experiments over benchmarks ranging from 0.51K to 13.16M non-linear activation units. Our evaluation shows that this novel technique can efficiently verify several formulae definable in the grammar and it outperforms ad-hoc encodings from the literature~\cite{casadio2022neural,athavale2024verifying}.
\end{itemize}

\noindent {\em Structure of the paper. } 
We define preliminaries in Section~\ref{sec:prelim}. Section~\ref{sec:examples} introduces different notions of confidence-based properties, the \softmax{} approximation, and the grammar. In Section~\ref{sec:framework}, we present our unifying technique to verify properties by encoding post-conditions as layers. 
Section~\ref{sec:experiments} reports our experiments demonstrating the effectiveness of our translation, and we conclude in Section~\ref{sec:conclusion}. Missing proof details and experiments can be found in the Appendix.

\paragraph{Related Work.}
In addition to the works already discussed, we discuss a few recent developments. The paper~\cite{athavale2024verifying} recently introduced softmax-based confidence and its verification focusing on computing bounds on the output of the softmax function based on the bounds of its input. Our approximation is different, as explained in Section~\ref{sec:relaxed_robust} (and App.~\ref{sec:analysis_compare}), and is fine-tunable with respect to the desired confidence level and usable with any neural network verified, not just constraint-based ones.
%
\cite{wei2023convex} introduces convex bounds on the softmax function, providing both lower and upper bounds that are compatible with convex optimization. 
A case-study surveying different robustness variants is presented in
\cite{casadio2022neural}
which includes strong and smooth robustness. The strong definition in~\cite{casadio2022neural} however, requires the logit value of the classified class should not drop below a certain threshold. 
We apply the same condition using confidence values, which are normalized between $0$ and $100$. 
Smoothness is an instance of Lipschitz continuity in neural networks~\cite{wang2022quantitative,rosca2020case,lindqvist2022novel,casadio2022neural,jordan2020exactly}, which bounds each class's logit values using a Lipschitz constant. Ours is a simplified version as we bound the confidence of the seed image's class by a threshold. 
The paper~\cite{naik2020robustness} performs robustness verification for cross-entropy loss functions, but there is a subtle difference, as we do not use the logarithm of the softmax output. Furthermore, the technique used in that paper is based on a {\em Satisfiability Modulo Convex Programming}~\cite{naik2020robustness} approach, whereas we use \alphabeta{}, which relies on a different suite of solver techniques, including PGD-attack~\cite{dong2018boosting}, CROWN~\cite{zhou2024scalable}, $\alpha$-CROWN~\cite{xu2020fast}, $\beta$-CROWN~\cite{wang2021beta}, among others. Other papers~\cite{GIUNCHIGLIA2024109124,shriver2021dnnv} have also explored encoding constraints into neural networks. The paper~\cite{GIUNCHIGLIA2024109124} appends a user-requirement as a layer to the neural network and trains the network to satisfy these requirements, however, their approach is limited to propositional formulae, and was not applied to post training robustness verification. The framework~\cite{shriver2021dnnv} simplifies neural networks by statically replacing unsupported operations with supported ones, enabling the underlying verifier to handle them. It also simplifies properties by decomposing them into smaller sub-properties that can be verified individually by the verifier. Finally,~\cite{NEURIPS2021_8df6a659} introduced the concepts of top-$k$ relaxed robustness and affinity robustness which we show can be captured in our approach. 

\section{Preliminaries}
\label{sec:prelim}

A neural network is composed of layers $l_0, l_1, \ldots, l_k$, where $k$ represents the number of layers. 
Each layer contains an ordered set of neurons. A neuron $n_{ij}$ represents the $j$-th neuron of layer $l_i$, 
and $|l_i|$ represents the number of neurons in layer $l_i$.  Layers $l_0$ and $l_k$ are the input and output layers, respectively; all other layers are hidden layers. We assume that input and output layers do not have any activation function, while each hidden layer $l_i$ has an activation function $\sigma_i$.
Some well-known non-linear activation functions include: \textsc{ReLU}: $\sigma_i(y) = \max(0, y)$; \textsc{LeakyReLU}: $\sigma_i(y) = \max(\alpha y, y)$, where $\alpha$ is the slope hyperparameter for the negative input region ($y \leq 0$); \textsc{Sigmoid}: $\sigma_i(y) = \frac{1}{1 + e^{-y}}$; and \textsc{Tanh}: $\sigma_i(y) = \frac{e^y - e^{-y}}{e^y + e^{-y}}$. We consider any activation function as long as it is supported by state of the art verifiers. 
Each layer $l_i$, except the input layer, has a weight matrix $W_i \in \mathbb{R}^{|l_i| \times |l_{i-1}|}$ and a bias vector $B_i \in \mathbb{R}^{|l_i|}$. Let $\Phi_i$ denote the function representing the computation from layer $l_0$ to $l_i$. Then, $\Phi_i(x) = \boldsymbol{\sigma}_i(W_i \Phi_{i-1}(x) + B_i)$, where $\boldsymbol{\sigma}_i$ denotes the vectorized version of the scalar activation function $\sigma_i$, and $\Phi_0(x) = x$. For the output layer $l_k$, the network output is given by $N(x) = \Phi_k(x) = W_k \Phi_{k-1}(x) + B_k$. $N_i(x)$ represents the logits value of $i^{th}$ output neuron. The final decision $\hat{N}$ of a neural network is determined using the \textsc{argmax}~\cite{bishop2006pattern} function, which returns the index/class of the maximum output value: $\hat{N}(x) = \textsc{argmax}(N(x))$. Thus, a neural network $N : \mathbb{R}^n \rightarrow \mathbb{R}^m$ is a function that takes an input of $n$ dimensions and produces an output of $m$ dimensions, i.e., $n = |l_0|$ and $m = |l_k|$. 


To construct a verification query, we define predicates $P$ and $Q$ over the input and output layers, respectively. A {\em verification query} is defined as a triple $\langle N, P, Q \rangle$. We say that the query holds if, for every input ${x}$ such that ${x} \models P$, it follows that $N({x}) \models Q$, i.e., $\forall x, P(x)\implies Q(N(x))$. If the query does not hold, then there exists a counterexample ${x}$ such that ${x} \models P$ and $N({x}) \not\models Q$.
For instance, the standard (local) robustness definition can be written as a verification query in the above form as follows.


\noindent{\it Standard Local Robustness.}
For a given image $\gi$ (sometimes called the seed image), a neural network $N$, and an input perturbation parameter $\epsilon$, the local robustness property~\cite{singh2019abstract,singh2019beyond,zhang2018efficient,xu2021fast,zhou2024scalable}, states that for all images $x$ close to $\gi$, the network's prediction for $x$ should be the same as its prediction for $\gi$. We represent the closeness as $\dist(\gi,x) = \|\gi - x\|_p$, where $p \in \{1,2,\infty\}$. Formally:
\begin{align}
    \label{eq:slr}
    \forall x \; \dist(\gi,x) \leq \epsilon \implies \hat{N}(\gi) = \hat{N}(x)
\end{align}
This corresponds to the verification query $\langle N,P,Q\rangle$, where predicate  $P$ is $\dist(\gi, x) \leq \epsilon$ and $Q$ is $\hat{N}(\gi) = \hat{N}(x)$. We call $x$ a {\em misclassified} input if $\hat{N}(\gi) \neq \hat{N}(x)$.  

\noindent{\it Softmax-based confidence computation in Neural Networks.}
In a neural network classifier, each output neuron is associated with a class, and the output of the neuron is the logit value for that class. The \softmax{} function converts these logits into a probability distribution over the classes, which indicates the confidence of the network in its prediction. This is formally defined in Eq~\ref{eq:softmax} below, where $t$ denotes the class for which we compute the confidence, and $y_i$ denotes the logit value for the $i^{th}$ dimension (class $i$) of the output layer. The tuple $(y_1,..,y_m)$ is denoted by $\bar{y}$. This definition is adopted from~\cite{ia2016deep,wei2023convex,athavale2024verifying}.
{\small
\begin{align}
    \label{eq:softmax}
\Conf((y_1,..,y_m),t) = \Conf(\bar{y},t) = 100\cdot\softmax(\bar{y},t) = \frac{100 \cdot e^{y_t}}{\sum_{i=1}^{m} e^{y_i} } .  
\end{align}
}

\section{Modeling Different Robustness Variants}
\label{sec:examples}
Different robustness properties are required for different applications~\cite{casadio2022neural} and hence it is important to allow users to specify and verify new properties as needed. At the same time incorporating confidence into properties poses an additional challenge (as also noted in~\cite{athavale2024verifying,casadio2022neural}), since confidence (as defined in Eq.~\ref{eq:softmax}) involves exponential functions. To integrate such properties with existing solver technologies, which primarily handle linear constraints or their Boolean combinations, one must approximate confidence in a suitable form. In this section, we present a grammar that captures a wide variety of robustness specifications, both with and without confidence. We then show our approximation of confidence and subsequently demonstrate how the confidence-based robustness variants discussed in the introduction are in this grammar.

%

\subsection{Generalized Reasoning on Confidence}
\label{sec:conf_reasoning}
The following grammar describes the rules for the post-conditions considered in our analysis. {\em LE} denotes linear expressions. {\em CC} represents confidence constraints, and the final property is represented by {\em PC} (for PostCond), which denotes Boolean combinations of linear constraints and {\em CC}. Let $\bar{y} = (y_1, y_2, \dots, y_m)$ represent the output logits of the neural network.
\[
\begin{aligned}
    \text{LE} &::= c_1y_1 + c_2y_2 + \dots + c_my_m + b, \forall i\in[m],  c_i,b \in \mathbb{R} \\
    \text{CC} &::= \Conf (\bar{y}, t) \bowtie b, t \in [m], b \in \mathbb{R^+}, \bowtie \in \{\leq,\geq,<,>\} \\
    \text{PC} &::= \text{PC} \land \text{PC} \mid \text{PC} \lor \text{PC} \mid \text{CC} \mid \text{LE} > 0 \mid \text{LE} \geq  0
\end{aligned}
\]
This grammar provides a unified representation for expressing a broad class of post-conditions that may combine logical constraints with confidence-based reasoning. Since the grammar involves expressions containing $\Conf$, our first step is to approximate $\Conf$ in a way that yields linear constraints over rationals, also called linear rational arithmetic (or LRA for short). It is essential that the approximation produces only LRA constraints, because in the next section, we demonstrate that any constraint expressible in LRA form can be automatically verified using our layer-based encoding.

\subsection{Approximation of $\Conf$}
As explained in Eq.~\ref{eq:softmax}, $\Conf$, is derived from the well-known \softmax{} function, which is highly non-linear, as it uses exponential terms. We approximate the $\Conf$ constraints by splitting into the following two cases:

{\noindent \textbf{Approximation of $\Conf (\bar{y}, t) \leq b$ and $\Conf (\bar{y}, t) < b$:}}
We consider the approximation of $\Conf(\bar{y}, t) < b$, and the approximation of $\Conf(\bar{y}, t) \leq b$ follows analogously. Recall that $\bar{y} = (y_1, \dots, y_m) = N(x)$. Suppose the dimension for which we analyze the confidence is $t$, and $y_t$ is the corresponding logit value. Consider the output dimension $t'$ such that the corresponding logit value is $y_{t'} = \max_{i=1,\, i \neq t}^{m} (y_i)$. For any $b > 0$, we have the following. Let $\delta = -\ln\!\left(\frac{100}{b} - 1\right)$.

\begin{customclaim}
  \label{claim:less_ineq}
  If $y_t < y_{t'}+\delta$ holds, then $\Conf(\bar{y}, t) < b$, else $\Conf (\bar{y}, t) \geq \frac{100}{1+(m-1)e^{-\delta}}$. 
\end{customclaim}
\begin{proof}
  If $y_t < y_{t'} + \delta$, by removing all exponential terms from the denominator except for $e^{y_t}$ and $e^{y_{t'}}$, we obtain 
  $
  \Conf(\bar{y}, t) \leq \frac{100e^{y_t}}{e^{y_{t}}+e^{y_{t'}}}.
  $
  After scaling the right-hand side by $\frac{1}{e^{y_t}}$, we get 
  $
  \Conf(\bar{y}, t) \leq \frac{100}{1+e^{-(y_t - y_{t'})}}.
  $ 
  Since $\delta > y_t- y_{t'}$, we have
  $
  \Conf(\bar{y}, t) < \frac{100}{1+e^{-\delta}}.
  $ 
  Since $\delta=-\ln(\frac{100}{b}-1)$,
  $
  \Conf(\bar{y}, t) < b.
  $ 
  Suppose the assumption does not hold, then 
  since $y_{t'}$ is the maximum among other logits, we can derive 
  $
  \Conf(\bar{y}, t) \geq \frac{100e^{y_t}}{e^{y_{t}}+(m-1)e^{y_{t'}}},
  $
  replacing $e^i$ by $e^{t'}$ for all $i,i\neq t$. 
  After scaling the fraction in the RHS by $e^{y_t}$, we obtain
  $
  \Conf(\bar{y}, t) \geq \frac{100}{1+(m-1)e^{y_{t'} - y_t}}.
  $
  Since $y_t \geq y_{t'} + \delta$, we obtain
  $
  \Conf(\bar{y}, t) \geq \frac{100}{1+(m-1)e^{-\delta}}$.
  \qed
\end{proof}

The first part of Claim~\ref{claim:less_ineq} ensures that $\Conf(\bar{y}, t) < b$ whenever $y_t < y_{t'} + \delta$ holds, establishing soundness, while the other provides the error bound, stating that if $y_t < y_{t'} + \delta$ does \emph{not} hold, then we characterize the lower bound of $\Conf(\bar{y}, t)$. The relationship between $\thresh$ and the error bound is illustrated in Figure~\ref{fig:lb-ub-approx} (App.~\ref{sec:analysis_compare}). The condition $y_t < y_{t'} + \delta$ can be encoded in LRA:
\begin{align}
  \label{eq:less:holds}
  \textstyle
  \bigvee_{i=1,\, i \neq t}^{m} \; y_t < y_i + \delta.
\end{align}
{\noindent \textbf{Approximation of $\Conf (\bar{y}, t) \geq b$ and $\Conf (\bar{y}, t) > b$:}}
Here also, we consider the approximation of $\Conf(\bar{y}, t) > b$, and the approximation of $\Conf(\bar{y}, t) \geq b$ follows analogously. Similar to the above approximations, let $t'$ be such that $y_{t'} = \max_{i=1,\, i \neq t}^{m} (y_i)$. Let $\delta = -\ln\!\left(\frac{1}{m-1}\left(\frac{100}{b}-1\right)\right)$. Then, for any $b > 0$, we have the following claim with proof in Appendix~\ref{sec:examples:proofs}.

\begin{customclaim}
  \label{claim:greater_ineq}
  If $y_t > y_{t'} + \delta$ holds, then $\Conf(\bar{y}, t) > b$, else $\Conf(\bar{y}, t) \leq \frac{100}{1+e^{-\delta}}$. 
\end{customclaim}

Again the first part of Claim~\ref{claim:greater_ineq} ensures soundness, and while the second provides the upper bound on the violation. Similar to Eq~\ref{eq:less:holds}, the condition $y_t > y_{t'} + \delta$ can also be written in LRA: $\bigwedge_{i=1,\, i \neq t}^{m} \; y_t > y_i + \delta.$

Suppose $CC^\#$ is the abstract constraint corresponding to a confidence constraint $CC$ in our grammar. For example, $y_t < y_{t'} + \delta$ is the abstract constraint for the confidence constraint $\Conf(\bar{y}, t) < b$ by Claim~\ref{claim:less_ineq}. By replacing $CC$ with $CC^\#$ in our grammar, we obtain the abstract post-condition $PC^\#$. The following theorem establishes soundness, with proof in Appendix~\ref{sec:examples:proofs}. Moreover, since \(CC^\#\) is in LRA form, it follows directly that \(PC^\#\) is also expressible in LRA.
\begin{theorem}
  \label{th:grammar:soundness}
  For any post-condition $PC$, the abstraction $PC^\#$ constitutes a sound abstraction of $PC$.
\end{theorem}

\subsection{Few Instances of Grammar}
We now show how the notions of relaxed robustness, strong robustness, and smoothness are in our grammar as specific practical instances. 

{\noindent \textbf{Relaxed Robustness:}}
\label{sec:relaxed_robust}
As discussed in Introduction, {\em relaxed robustness}, is a weaker notion where low-confidence counterexamples are ignored in determining robustness of the network, as shown in Figure~\ref{fig:mot:combine}(a-b). For all images $x$ close to given image $\gi$, their classification should be the same, unless the confidence of the network on $x$ is less than a certain threshold. Incorporating the requirement in Eq~\ref{eq:slr} gives us Eq~\ref{eq:rr} which permits low-confidence counterexamples to bypass the strict robustness criteria. Let $\thresh$ be the confidence level threshold, a tunable user specified parameter.
\begin{align}
    \label{eq:rr}
    \forall x \; & \dist(\gi,x) \leq \epsilon
    \implies \left(\Conf(N(x),\hat{N}(x)) < \thresh \vee \hat{N}(\gi) = \hat{N}(x) \right)
\end{align}
Then in the verification query $\langle N,P,Q\rangle$, $N$ and $P$ remain same as in Eq~\ref{eq:slr}, while the post-condition $Q$ now gets changed to $Q_{rel}=\Conf(N(x),\hat{N}(x)) < \thresh \vee \hat{N}(\gi) = \hat{N}(x)$. Letting $\bar{y} = N(x)$, $t=\hat{N}(x)$, $c=\hat{N}(\gi)$, we can rewrite $Q_{rel}$ as: $\Conf(\bar{y},t) < \thresh \vee \bigwedge_{i=1,i \neq c}^{m} y_c > y_i$, which is in our grammar.
%
Applying Theorem~\ref{th:grammar:soundness} we get an approximation of $Q_{rel}$, which is $y_t < y_{t'} + \delta \vee \bigwedge_{j=1,j \neq t}^{m} y_c > y_j$, where $y_{t'}$ and $\delta$ are the same as in Claim~\ref{claim:less_ineq}.
Further, by replacing $y_t < y_{t'} + \delta$ from Eq~\ref{eq:less:holds}, we get $\bigvee_{j=1,j\neq t}^{m} y_t < y_i + \delta \vee \bigwedge_{j=1,j \neq c}^{m} y_c > y_j$. Since $t=\hat{N}(x)$ means $y_t = \max_{i=1}^{m} (y_i)$, to encode the constraints for $y_t$, we just compare $y_t$ with every other logits value, which finally gives:
\begin{align}
  \label{eq:rr:derived:reduced}
  \textstyle
    Q_{rel}' = (\bigwedge_{i=1}^{m} \bigvee_{j=1,j\neq i}^{m}  y_i < y_j + \delta) \lor \bigwedge_{i=1,i\neq c}^{m} y_i < y_c
\end{align}
Theorem~\ref{th:grammar:soundness} ensures soundness for $Q_{rel}'$ i.e., $\langle N, P, Q'_{rel} \rangle \implies \langle N, P, Q_{rel}\rangle$. In fact, from Claim~\ref{claim:less_ineq} we can also obtain a lower bound on the confidence of any counterexample as $\frac{100}{1 + (m-1)e^{ - \delta}}$.
{\noindent \textbf{Strong Robustness:}}
The second variant of robustness, that we call {\em strong robustness}, aims to capture cases where the confidence on the seed image is high, but drops below a certain threshold within an $\epsilon$-bounded perturbation, regardless of whether the classification label changes. In other words, a significant drop in confidence itself indicates a weakness in the underlying network. Recall that an example of such robustness was illustrated in Figure~\ref{fig:mot:combine}(c-e), where the confidence for a seed image of class \texttt{ship} dropped from $96.17\%$ to $22.21\%$. The definition of strong robustness, which generalizes the notion introduced in~\cite{casadio2022neural}, can now be formally defined as follows. Let $\thresh_1 > \thresh_2$ be two threshold values.
\vspace{-2mm}
\begin{align}
    \label{eq:sr}
     \forall  x \; \dist(\gi,x) \leq \epsilon  \implies &\left(\Conf(N(x),\hat{N}(\gi)) > \thresh_2 \land \hat{N}(\gi) = \hat{N}(x)\right)\notag\\ & \lor \Conf(N(\gi),\hat{N}(\gi)) < \thresh_1 
\end{align}
The above equation asserts that for any $x$ in the $\epsilon$-neighborhood of $\gi$, if confidence of the network on $\gi$ is at least $\thresh_1$, then $x$ must be classified the same as $\gi$ with a high confidence (above threshold $\thresh_2$). Thus, we get the query $\langle N,P,Q_{str}\rangle$ where $P$ remains the same, while $Q_{str}=(\Conf(N(x),\hat{N}(\gi)) > \thresh_2 \land \hat{N}(\gi) = \hat{N}(x)) \lor \Conf(N(\gi),\hat{N}(\gi)) < \thresh_1$. Note that $\Conf(N(\gi),\hat{N}(\gi)) < \thresh_1$ can be computed beforehand since $\gi$ is a given seed image. Thus, we will only need to show the condition $\Conf(N(x),\hat{N}(\gi)) > \thresh_2 \land \hat{N}(\gi) = \hat{N}(x)$. Let $t = \hat{N}(\gi)$ and $(y_1,..,y_m) = N(x)$. The condition $\hat{N}(\gi) = \hat{N}(x)$ can be represented as $\bigwedge_{i=1,i\neq t}^{m} y_t > y_i$, so, $Q_{str}$ can be written as $\Conf(N(x),t) > \thresh_2 \land \bigwedge_{i=1,i\neq t}^{m} y_t > y_i$, which is in our grammar. 

{\noindent \textbf{Smoothness:}}
A third variant is smoothness, an instance of Lipschitz continuity in neural networks~\cite{wang2022quantitative,rosca2020case,lindqvist2022novel}. The work in~\cite{casadio2022neural} introduces the concept of Lipschitz robustness, which bounds each class's logit values using a Lipschitz constant. We propose a simplified instance of Lipschitz robustness, where we ask that, within an $\epsilon$-perturbation, the confidence of network should not exhibit significant variations wrt the seed image. 
Let $t = \hat{N}(\gi)$, and $\thresh>0$ a threshold, then: \begin{align}
  \forall x \;  \dist(\gi,x) \leq \epsilon \implies 
  | \Conf(N(\gi),t) - \Conf(N(x),t) | < \thresh
\end{align}

As before, $\Conf(N(\gi), t)$ can be pre-computed as $\gi$ is a given seed image; so we can fix $\Conf(N(\gi), t) = C$, a constant. Substituting this, we obtain a verification query $\langle N, P,Q_{sm}\rangle$, where $Q_{sm}= C - \Conf(N(x), t) > -\thresh \land C - \Conf(N(x), t) < \thresh$, with $P$ unchanged. $Q_{sm}$ can be further simplified to $\Conf(N(x),t) < C + \thresh \land \Conf(N(x),t) > C - \thresh$, here $C+\thresh$ and $C-\thresh$ are constants. This simplified post-condition is again in our grammar. 

{\noindent \textbf{Top-k robustness:}}
Till now, we have discussed robustness variants involving confidence, now we show that even other variants are also instances of our grammar. We consider the notion of top-$k$ robustness introduced in~\cite{NEURIPS2021_8df6a659}. Let $N(x,k)$ denote the $k$-th highest logit value of the network output, and define $N^k(x) = \{i \mid N_i(x) \ge N(x,k)\}$, the set of classes with the top-$k$ highest logits. Top-$k$ robustness is then defined as $\forall x'\;\mathrm{dist}(x,x') \le \epsilon \;\implies\; N^k(x) = N^k(x')$. See Figure~\ref{fig:topk} in Appendix~\ref{app:topk} for an example. The same paper~\cite{NEURIPS2021_8df6a659} also defines a relaxed variant (\emph{top-$k$-relaxed}) robustness: $\forall x'\;\mathrm{dist}(x,x') \le \epsilon \;\implies\; \exists k \le K\;:\;N^k(x) = N^k(x')$. Another variant, \emph{top-$k$-affinity}, permits the user to specify the set of classes in which misclassification is allowed. All three variants of top-$k$ robustness are captured by our grammar (see Appendix~\ref{app:topk}).

\section{Encoding Mechanism via Additional Layers}
\label{sec:framework}

In this section, we provide an encoding for all robustness variants definable by the grammar, such that they can be integrated in state-of-the-art robustness verification engines. We begin by noting that the International Verification of Neural Networks Competition (\vnncomp{})~\cite{brix2024fifth} standardizes both the neural network format (ONNX) and the property format (\vnnlib{}). All neural network verifiers participating in \vnncomp{} take as input a neural network in ONNX format and a property file in \vnnlib{} format. Importantly, the \vnnlib{} format encodes pre-condition $P$ and negation of the post-condition $Q$ of the neural network $N$, expressed as an arbitrary Boolean combination of linear constraints, i.e., exactly in the grammar that we described in the previous section. However, most state-of-the-art neural network verifiers are optimized for {\em simplified post-conditions}, typically involving only disjunctions or conjunctions of linear atoms over the outputs of the network.
In what follows, we call a post-condition {\em simplified}  if it is either atomic $y > 0$, $y \geq 0$, $y < 0$, or $y \leq 0$, where $y$ is one of the outputs of $N$ or conjunction/disjunction of the atomics.
  
In this section, we show how we can convert post-conditions into simplified ones {\em by just appending a few additional layers} to the neural network, without having to change the encoding within the verifiers. In doing so, we only add a linear number of neurons in the size of the post-condition and number of the layers added is at most the depth of the formula (represented as a Directed Acyclic Graph) in the post-condition. As we will see, while it is easy to do this for conjunctions, it is non-trivial when both conjunctions and disjunctions are present, and we achieve a solution with an approximation factor.

\noindent{\it Translating Conjunctions.}
Translating conjunctions is relatively simple (as also done in~\cite{shriver2021dnnv}). 
Given $\Land_{i=0}^n LE_i \leq 0$, where nodes $y_1, \dots , y_m$ represent the original network's output nodes and $LE_i = \sum_{j=1}^{m} w_{ij}y_j+b_j$, defining a new output $y = \sum_{i=0}^{n} \relu{}(LE_i)$ can be encoded as an additional layer as depicted in Figure~\ref{fig:conj}(a). Nodes $LE_1, LE_2, \dots, LE_n$ represent actions of corresponding linear expressions followed by \relu{} nodes. We immediately obtain (proof in App.~\ref{sec:framework-proofs}):
%

\begin{lemma}
  \label{thm:encoding_conj}
  $y \leq 0 \Longleftrightarrow \Land_{i=0}^n LE_i \leq 0$
\end{lemma}
%

%
%
The number of \relu{} nodes added is the number of clauses in the property, e.g., in Figure~\ref{fig:conj}(a), there are a total of $n$ clauses; therefore, $n$ \relu{} nodes are added. We also deduce that $y > 0 \Leftrightarrow \Lor_{i=0}^n LE_i > 0$, which encodes disjunction. \medskip

\noindent{\it The general translation.}
When we have both disjunctions and conjunctions in the post-condition, we cannot directly feed the output of conjunctions to the input of a disjunction, since 
for conjunction, $y = 0$ is interpreted as true and $y > 0$ as false, while for disjunction, the roles are reversed. 
A straightforward approach to handle this asymmetry is to convert the post-condition into Disjunctive Normal Form (DNF) and execute each clause in parallel.  However, this method suffers from several drawbacks: (i) the conversion to DNF may cause exponential blowup, (ii) information derived from one clause cannot be exploited by another, and (iii) the underlying verifier codebase may still need modifications to accommodate each clause. Another way to encode disjunctions is by using products, however, this introduces highly non-linear constraints, which are very inefficient to handle. Although some post conditions are expressible in DNF form, this is not always the case. For instance, for the top-k properties introduced in Section~\ref{sec:examples} the post-conditions are not in DNF (also see Appendix~\ref{app:topk} Equations~\ref{app:eq:topk:rel} and \ref{app:eq:topk:aff})

\newcommand{\ckt}{\mathfrak{C}}

Therefore, we take an alternate approach via a transformation $flip(b,v) = b-v$ of a signal $v$ for some $b>0$ (defined later), which allows us to feed output of disjunction to conjunction and vice-versa. Our translation will also introduce non-linear constraints but only as $Relu$'s which can be easily modeled in verifiers.
%
In the following, let $\dagger \in \{\lor,\land\}$ and if $\dagger = \lor$, then $\bar{\dagger} = \land$, if $\dagger = \land$, then $\bar{\dagger} = \lor$. Wlog., we also assume that the post-condition is given in Negation Normal Form, with negations pushed inside the linear inequalities, with $\land$ and $\lor$ flattened (i.e., $\land$ and $\lor$ alternate when traversing the formula top-down). Then, $\dagger\in \{\lor,\land\}$, for any post-condition $Q\in \mathcal{PC}$ and a parameter $\eta\in\mathbb{R}$, we define the gadget $V(\dagger,Q,\eta)$ as an expression made up of $flip$s, $Relu$, and summation functions. Hence for every such expression we can build a circuit $\ckt_{V(\dagger,Q,\eta)}$ that can be integrated with the Neural network using additional layers (for $flip$ we use $-1$ weight on edges, while $Relu$ and summation are standard operators in neural networks). Thus, given a set of input values to $Q$, $V(\dagger,Q,\eta)$ evaluates to a real output value $y$, satisfying some properties, as we shall describe below. Both $V$ and its circuit are defined inductively based on the structure of the post-condition.
%
  For atomic formulas, we construct the gadgets $V$ as given below, one for each value of $\dagger\in\{\land,\lor\}$:
  \begin{align}
    \label{eq:base_gadget}
    V(\land, LE\leq 0,\smallparam) = LE+\smallparam \qquad 
    V(\lor, LE \leq 0,\smallparam) = -LE+\smallparam 
  \end{align}
The following gives the gadget for $Q= Q_1 \dagger .. \dagger Q_k$ (where $\dagger\in\{\lor,\land\}$), whose circuit is pictorially illustrated in Figure~\ref{fig:conj}(b).
\begin{align}
  \label{eq:generic_gadget}
  \textstyle  
  V(\dagger, Q,\smallparam) &= \sum\nolimits_{i=1}^{k}Relu(flip(b(\dagger,k,\smallparam), V(\bar{\dagger},Q_i,\smallparam)))\\
  \label{eq:def_b}
    &\text{where,} b(\dagger,k,\smallparam) = \begin{cases} \smallparam(1+1/k) & \mbox{if } \dagger = \land \\ 2*\smallparam  & \mbox{if } \dagger = \lor \end{cases}
\end{align}
The main intuition of the above construction is that after crossing every disjunct/conjunct, the direction of the bounds flip, i.e., upper bound on an output becomes a lower bound and vice versa. This allows us to propagate bounds when the circuit has disjuncts and conjuncts. Before formalizing this intuition and proving the properties of the translation, we present in Figure~\ref{fig:dnf}(c), an illustrative example. In this example, the negation of the post-condition $\neg Q$ is $((y_1 + y_2 \leq 0 \land y_2 \leq 0) \lor (y_1 - y_3 \leq 0 \land y_3 \leq 2))$, with atomic formulae $y_1 + y_2 \leq 0$, $y_2 \leq 0$, and so on. The parameter $\smallparam$ in the input constraint is set to $0.2$ (user-defined). The part of circuit before the first dashed vertical line are the gadgets for atomic formulae. For instance, $y_1 + y_2 \leq 0$ is converted into $-y_1 - y_2 + 0.2$, which is represented by the red part of the circuit. 
Thus, the output at the red line satisfies $0.2\leq -y_1-y_2+0.2$ which is equivalent to $y_1+y_2\leq 0$. The reason to have this flipped representation is that $y_1+y_2\leq 0$ and $y_2\leq 0$ are connected with conjunction, but the formulation of Eq~\ref{eq:generic_gadget} expects the atomic formulae to be connected with disjunction, thus we apply the disjunctive rule in right had side of of \eqref{eq:base_gadget}. The resulting outputs are connected with conjunctions, and the gadget between the two vertical lines corresponds to these conjunctions. 
Since this gadget represents conjunction, the value of $b$ is determined by the conjunction case in Eq~\ref{eq:def_b}, i.e., $b = \smallparam \left(1 + \tfrac{1}{k}\right) = 0.2 \left(1 + \tfrac{1}{2}\right) = 0.3$. In other words, at the blue line the output will be less than or equal to $0.2$ if $(y_1 + y_2 \leq 0 \land y_2 \leq 0)$ (as shown more generally in Lemma~\ref{thm:ub3} below). Finally, the conjunctions are connected with disjunctions, represented by the gadget after the second vertical line, which is also constructed following Eq~\ref{eq:def_b}, where $b = 2\smallparam = 0.2 \times 2 = 0.4$.


\begin{figure*}[t]
  \centering
  \begin{minipage}{0.4\linewidth}
    \centering
    \scalebox{0.8}{
      \begin{tikzpicture}[node distance = 11mm,scale=0.8]
    \node[] (y1)    at (0,0mm) {$y_1$};
    \node[below of=y1,yshift=0.5cm] (y2) {$y_2$};
    \node[below of=y2,yshift=0.9cm] (y3) {.};
    \node[below of=y3,yshift=0.9cm] (y4) {.};
    \node[below of=y4,yshift=0.9cm] (y5) {.};
    \node[below of=y5,yshift=6mm] (ym) {$y_m$};

    \node[draw,dotted,left of = y1,xshift=2mm,yshift=-8mm,minimum height=20mm] (N)  {$N$};
    \draw[->,thick] ($(y1) + (-8mm,0)$) -- (y1);
    \draw[->,thick] ($(y2) + (-8mm,0)$) -- (y2);
    \draw[->,thick] ($(ym) + (-8mm,0)$) -- (ym);

    \node[left of = N] (x)  {$x$};
    \draw[->,thick] (x) -- (N);

    \node[draw,below right of = y1,xshift=6mm,yshift=6mm] (le1)  {$b_1$};
    \node[below of=le1,yshift=0.6cm] (le3) {.};
    \node[below of=le3,yshift=0.9cm] (le4) {.};
    \node[draw,below of=le4,yshift=0.5cm] (len) {$b_n$};
    \draw[->,thick] (y1) -- node[above]  {$c_{11}$} (le1);
    \draw[->,thick] (y2) -- node[above]  {$c_{12}$} (le1);
    \draw[->,thick] (ym) --  (le1);


    \draw[->,thick] (y1) --  (len);
    \draw[->,thick] (y2) --  (len);
    \draw[->,thick] (ym) -- node[below]  {$c_{nm}$} (len);

    \node[draw,circle,right of = le1,xshift=-3mm] (r1) {};
    \node[draw,circle,right of=len,xshift=-3mm] (rn) {};

    \draw[->,thick] (le1) -- (r1);
    \draw[->,thick] (len) -- (rn);

    \node[draw,right of = r1,xshift=-3mm,yshift=-6.5mm] (a1) {0};

    \draw[->,thick] (r1) -- node[above]  {$1$} (a1);
    \draw[->,thick] (rn) -- node[below]  {$1$} (a1);

    \node[right of =a1,xshift=-3mm] (c) {$y$};
    \draw[->,thick] (a1) -- (c);

  \end{tikzpicture}
  
      }
     \\
  \vspace{-4mm}\hspace{3cm} (a)\par
  \vspace{3mm}
    \scalebox{0.8}{
      \begin{tikzpicture}[node distance = 11mm,scale=0.8]
    \node[draw,dotted] (y1)    at (0,0mm) {$\ckt_{V(\bar{\dagger},Q_1,\smallparam)}$};
    \node[draw,dotted,below of=y1,yshift=-0.2cm] (y3) {$\ckt_{V(\bar{\dagger},Q_k,\smallparam)}$};

    \node[draw,right of = y1, xshift=12mm] (le1)  {$b(\dagger,k,\smallparam)$};
    \node[below of=le1,yshift=0.6cm] (le3) {.};
    \node[below of=le3,yshift=0.9cm] (le4) {.};
    \node[draw,below of=le4,yshift=0.5cm] (len) {$b(\dagger,k,\smallparam)$};

    \draw[->,thick] (y1) -- node[above] {$-1$} (le1);
    \draw[->,thick] (y3) -- node[above] {$-1$} (len);

    \node[draw,circle,right of = le1] (r1) {};
    \node[draw,circle,right of=len] (rn) {};

    \draw[->,thick] (le1) -- (r1);
    \draw[->,thick] (len) -- (rn);

    \node[draw,right of = r1, yshift=-6.5mm,xshift=-4mm] (a1) {0};

    \draw[->,thick] (r1) -- node[above]  {$1$} (a1);
    \draw[->,thick] (rn) -- node[below]  {$1$} (a1);

    \node[right of =a1,xshift=-2mm] (c) {}; 
    \draw[->,thick] (a1) -- (c);

  \end{tikzpicture}
  
      }
      \\
  \vspace{-3mm}\hspace{3cm} (b)
  \end{minipage}
  \begin{minipage}{0.59\linewidth}
  \centering
   \scalebox{0.8}{
    \begin{tikzpicture}[node distance = 11mm,scale=0.8]
    \node[] (y1)    at (0,0mm) {$y_1$};
    \node[below of=y1] (y2) {$y_2$};
    \node[below of=y2] (y3) {$y_3$};

    \node[draw,above right of = y1,xshift=6mm,yshift=-2mm,red] (le1)  {0.2};
    \node[draw,below of=le1] (le2) {0.2};
    \node[draw,below of=le2] (le3) {0.2};
    \node[draw,below of=le3] (le4) {2.2};
    \draw[->,thick,red] (y1) -- node[above]  {$-1$} (le1);
    \draw[->,thick,red] (y2) -- node[above]  {$-1$} (le1);
    \draw[->,thick] (y2) -- node[above]  {$-1$} (le2);
    \draw[->,thick] (y1) -- node[below]  {$-1$} (le3);
    \draw[->,thick] (y3) -- node[above]  {$1$} (le3);
    \draw[->,thick] (y3) -- node[above]  {$-1$} (le4);

    \node[below of=le4,xshift=-5mm] (text) {Base};

    \node[draw,right of=le1, xshift=4mm] (fle1)  {0.3};
    \node[draw,below of=fle1] (fle2) {0.3};
    \node[draw,below of=fle2] (fle3) {0.3};
    \node[draw,below of=fle3] (fle4) {0.3};
    
    \draw[->,thick,red] (le1) -- node[above,xshift=1mm] {$-1$} (fle1);
    \draw[->,thick] (le2) -- node[above,xshift=1mm] {$-1$} (fle2);
    \draw[->,thick] (le3) -- node[above,xshift=1mm] {$-1$} (fle3);
    \draw[->,thick] (le4) -- node[above,xshift=1mm] {$-1$} (fle4);
    
    \node[draw,circle,right of = fle1,xshift=-3mm] (r1) {};
    \node[draw,circle,below of=r1] (r2) {};
    \node[draw,circle,below of=r2] (r3) {};
    \node[draw,circle,below of=r3] (r4) {};

    \node[below of=r4,xshift=-5mm] (text) {Conjunctions};

    \draw[->,thick] (fle1) -- (r1);
    \draw[->,thick] (fle2) -- (r2);
    \draw[->,thick] (fle3) -- (r3);
    \draw[->,thick] (fle4) -- (r4);

    \node[draw,below right of = r1,yshift=2mm] (a1) {0};
    \node[draw,below right of = r3,yshift=2mm] (a2) {0};

    \draw[->,thick] (r1) -- node[above]  {$1$} (a1);
    \draw[->,thick] (r2) -- node[above]  {$1$} (a1);
    \draw[->,thick] (r3) -- node[above]  {$1$} (a2);
    \draw[->,thick] (r4) -- node[above]  {$1$} (a2);

    \node[draw,right of = a1,xshift=2mm] (m1) {0.4};
    \node[draw,right of = a2,xshift=2mm] (m2) {0.4};
    \draw[->,thick,blue] (a1) -- node[above,xshift=1mm]  {$-1$} (m1);
    \draw[->,thick] (a2) -- node[above,xshift=1mm]  {$-1$} (m2);

    \node[draw,circle,right of = m1,xshift=-4mm] (rr1) {};
    \node[draw,circle,right of = m2,xshift=-4mm] (rr2) {};
    \draw[->,thick] (m1) -- (rr1);
    \draw[->,thick] (m2) -- (rr2);

    \node[draw,below right of = rr1,yshift=-4mm,xshift=-3mm] (f) {0};
    \draw[->,thick] (rr1) -- node[right]  {$1$} (f);
    \draw[->,thick] (rr2) -- node[right]  {$1$} (f);

    \node[right of =f,xshift=1mm] (c) {$y \geq 0.2 $};
    \draw[->,thick] (f) -- (c);

    \node[below of=r4,xshift=24mm] (text) {Disjunction};

    \path (le1) -- (fle1) coordinate[midway] (midline);

    \draw[dashed] ([xshift=-3mm, yshift=6mm]midline |- le1.north) -- ([xshift=-3mm, yshift=-6mm]midline |- le4.south);

    \path (a1) -- (m1) coordinate[midway] (midline1);

    \draw[dashed] ([xshift=-2mm, yshift=12.8mm]midline1 |- a1.north) -- ([xshift=-2mm, yshift=-12.8mm]midline1 |- m2.south);

  \end{tikzpicture}
   }  \\
  (c)
 \end{minipage}
  \caption{
    (a) Neural network $N$ appended with neural network layer that encodes either $\Land_{i=0}^n LE_i \leq 0$ and its negation $\Lor_{i=0}^n LE_i > 0$, where $LE_i = \sum_{j=1}^{m} c_{ij}y_j+b_i$. The circular nodes are \relu{} and the square nodes nodes are linear combinations.
    (b) The circuit for $\ckt_{V(\dagger,Q,\smallparam)}$ 
    (c) Translation of post-condition $\lnot ((y_1 + y_2 \leq 0 \land y_2 \leq 0 ) \lor (y_1 - y_3 \leq 0 \land y_3 \leq 2 ))$ using our scheme and $\smallparam = 0.2$.
  }
  \label{fig:conj}
  \label{fig:dnf}
\end{figure*}

Our choice of $b(\dagger,k,\smallparam)$ is crucial for ensuring the desired properties of the gadget. The following technical lemma (with proof in Appendix~\ref{sec:framework-proofs}) establishes how can we propagate the bounds on the signal $V(\dagger, Q,\smallparam)$ to the sub-formulas $Q_i$ and vice-versa, which will help us relating bounds on the inputs of the gadget and the bounds on the output of the gadget.
%
%
\begin{lemma}For $\smallparam > 0$ and $\beta > 0$, the following holds for $Q$.
    \label{thm:ub3}
  \begin{enumerate}
\item  
  $\forall i.\;\smallparam \leq V(\lor,Q_i,\smallparam) \limplies V(\land,Q,\smallparam) \leq \smallparam $.
\item   
  $\exists i.\; V({\land},Q_i,\smallparam) \leq \smallparam \limplies \smallparam \leq V(\lor,Q,\smallparam)$.
\end{enumerate}
\end{lemma}
  Using the above lemma, we are now ready to state Lemma~\ref{lemm:full} (with proof in Appendix~\ref{sec:framework-proofs}) that establishes the correctness of our translation scheme.  
  The first two cases demonstrate correctness for atomic formulae, while Cases~3 and~4 establish correctness for general formulae. 
%
\begin{lemma}
  \label{lemm:full}
  For a given post condition $Q$ and an $\smallparam> 0$, the following holds
\end{lemma}
    \begin{enumerate}
  \item If $Q = LE \leq 0$ , $V(\land, Q, \smallparam) \leq \smallparam \lequiv Q$  
  \item If $Q = LE \leq 0$ , $V(\lor,  Q, \smallparam) \geq \smallparam \lequiv Q$
  \item If $Q$ is conjunctive, i.e., $Q=Q_1\land ... \land Q_k$, then $Q \limplies V(\land, Q,\smallparam) \leq \smallparam$
  \item If $Q$ is disjunctive, i.e., $Q=Q_1\lor ... \lor Q_k$, then $Q \limplies V(\lor, Q,\smallparam) \geq \smallparam$
  \end{enumerate}
    Using the above, we can finally define $N' = Append(N, V(\dagger, Q, \smallparam))$ to be the neural network obtained by attaching the circuit  $\ckt_{V(\dagger, Q, \smallparam)}$ after the neural network $N$ with $y$ being the final output.
 More precisely, the inputs of $V(\dagger, Q, \smallparam)$ are connected to the outputs of $N$ and the output $y$ of $V(\dagger, Q, \smallparam)$ is the only output of $N'$. Finally, from the above lemma, we can prove our theorem (with proof in Appendix~\ref{sec:framework-proofs}) which justifies using it as a verification query. 
  \begin{theorem}
    \label{thm:query}
    Consider neural network $N$, pre-condition $P$, a post-condition $Q$,
    and $\smallparam > 0$. Let $Z$ be the negation normal form of $\lnot Q$.
    Let $N' = Append(N, V(\dagger, Z, \smallparam))$ be the neural network obtained by
    attaching $\ckt_{V(\dagger, Q, \smallparam)}$ after neural network $N$ and $y$ be the final output.
    \begin{enumerate}
      \item If the top symbol of $Z$ is disjunction, $\langle N', P, y < \smallparam \rangle \limplies \langle N, P, Q \rangle$
      \item If the top symbol of $Z$ is conjunction, $ \langle N', P, y > \smallparam \rangle \limplies \langle N, P, Q \rangle$
    \end{enumerate}
  \end{theorem}

\paragraph{Integrating All Components: Soundness.}
We establish the soundness of our framework by integrating the two main components: confidence approximation (from Section~\ref{sec:examples}) and the encoding mechanism (as defined in this section). 

\begin{theorem}
  \label{thm:final}
  Let $N$ be a neural network, $P$ a pre-condition, $Q$ any post-condition satisfied by grammar (specified in Section~\ref{sec:examples}). Let $N'$ be the appended neural network and $Q'$ be the simplified post-condition obtained through the encoding (specified in this section). If the query $\langle N', P,  Q' \rangle$ holds, then the original query $\langle N, P, Q \rangle$ also holds.
\end{theorem}


In Appendix~\ref{sec:framework-proofs}, we provide its proof and also provide bounds on the error in the counterexamples incurred due to our encoding. Interestingly, we did not observe any counterexamples with errors in our experiments.
For simplicity of the presentation, we only considered non-strict inequalities above. In Appendix~\ref{sec:support-strict}, we provide versions of the theorems that handle strict inequalities.





\section{Experiments}
\label{sec:experiments}
Our implementation encodes the post-conditions into additional neural network layers as described in the previous section and then invokes a Neural network verification engine. 
Our experiments are designed to address the following research questions: 
\textbf{RQ1:} How does our approach compare with an ad-hoc encoding of properties in a \emph{constraint-based state-of-the-art solver}, e.g., \marabou{}?  
\textbf{RQ2:} 
  Does our layer-based approach work and scale for different variants of robustness on large \vnncomp{} benchmarks, as it enables us to use \alphabeta{}, which has consistently ranked $1^{\text{st}}$ in \vnncomp{} 2021--2024, on top of Marabou.
To demonstrate the empirical impact and meaningfulness of the different robustness variants introduced in Section~\ref{sec:examples}, we provide illustrative images in Appendix~\ref{app:topk} and Appendix \ref{app:exp}.

\begin{figure}[t]
  \centering
  \begin{subfigure}[t]{0.33\textwidth}
    \centering
    \begin{adjustbox}{width=\linewidth, height=0.3\textheight, keepaspectratio}
      \begin{tikzpicture}
    \begin{axis}[
        xlabel= {\LARGE Confidences},
        ylabel={\LARGE Percentage (timeout)},
        ylabel style={at={(axis description cs:0.07,0.5)}, anchor=south},
        width=11cm,
        height=8cm,
        xmin=55, xmax=100,
        ymin=0, ymax=100,
        xtick={55, 60, 80, 90, 95},
        ytick={0,20,40,60,80,100},
        tick label style={font=\LARGE},
        legend pos=north west,
        legend entries={\large \textsc{Marabou ad-hoc encod - MNIST}, \large\textsc{Marabou layered encod - MNIST}, \large\textsc{\alphabeta{} layered encod - MNIST}, \large\textsc{\alphabeta{} layered encod - others}},
        ymajorgrids=true,
        grid style=dashed,
    ]

    \addplot[
        color=cyan,
        mark=*,
    ]
    coordinates {
        (60,56.67)(80,48.89)(90,43.33)(95, 35.56)
    };

    \addplot[
        color=green,
        mark=*,
        dashed,
    ]     
    coordinates {
        (60,48.89)(80,35.56)(90,31.11)(95,30.0)
    };

    \addplot[
        color=black,
        mark=*,
        dashdotted,
    ]
    coordinates {
        (60,11.11)(80,4.44)(90,0)(95,0)
    };

    \addplot[
        color=red,
        mark=*,
        dash pattern=on 3pt off 2pt on 1pt off 2pt,
    ]
    coordinates {
        (60,11.5)(80,5)(90,4.5)(95,4.5)
    };

    \end{axis}

\end{tikzpicture}
    \end{adjustbox}
    \caption{Relaxed robust}
    \label{plot:compare-relaxed}
  \end{subfigure}%
  \begin{subfigure}[t]{0.33\textwidth}
    \centering
    \begin{adjustbox}{width=\linewidth, height=0.3\textheight, keepaspectratio}
      \begin{tikzpicture}
    \begin{axis}[
       xlabel= {\LARGE Confidences},
        ylabel={\LARGE Percentage (timeout)},
        ylabel style={at={(axis description cs:0.07,0.5)}, anchor=south},
        width=11cm,
        height=8cm,
        xmin=14, xmax=22,
        ymin=0, ymax=100,
        xtick={15, 17, 20},
        ytick={0,20,40,60,80,100},
        tick label style={font=\LARGE},
        legend pos=north west,
        legend entries={\large \textsc{Marabou ad-hoc encod - MNIST}, \large\textsc{Marabou layered encod - MNIST}, \large\textsc{\alphabeta{} layered encod - MNIST}, \large\textsc{\alphabeta{} layered encod - others}},
        ymajorgrids=true,
        grid style=dashed,
    ]

    \addplot[
        color=cyan,
        mark=*,
    ]
    coordinates {
        (15,30.2)(17,30.6)(20,31.2)
    };

    \addplot[
        color=green,
        mark=*,
        dashed,
    ]     
    coordinates {
        (15,21.2)(17,21.4)(20,18.6)
    };

    \addplot[
        color=black,
        mark=*,
        dashdotted,
    ]
    coordinates {
        (15,0)(17,0)(20,0)
    };

     \addplot[
        color=red,
        mark=*,
        dash pattern=on 3pt off 2pt on 1pt off 2pt,
    ]
    coordinates {
        (15,13)(17,12)(20,10)
    };

    \end{axis}

\end{tikzpicture}
    \end{adjustbox}
    \caption{Strong robust}
    \label{plot:compare-strong}
  \end{subfigure}
   \begin{subfigure}[t]{0.33\textwidth}
    \centering
    \begin{adjustbox}{width=\linewidth, height=0.3\textheight, keepaspectratio}
      \begin{tikzpicture}
    \begin{axis}[
        xlabel= {\LARGE Confidences},
        ylabel={\LARGE Percentage (timeout)},
        ylabel style={at={(axis description cs:0.07,0.5)}, anchor=south},
        width=11cm,
        height=8cm,
        xmin=5, xmax=45,
        ymin=0, ymax=100,
        xtick={10,25,40},
        ytick={0,20,40,60,80,100},
        tick label style={font=\LARGE},
        legend pos=north west,
        legend entries={\large \textsc{Marabou ad-hoc encod - MNIST}, \large\textsc{Marabou layered encod - MNIST}, \large\textsc{\alphabeta{} layered encod - MNIST}, \large\textsc{\alphabeta{} layered encod - others}},
        ymajorgrids=true,
        grid style=dashed,
    ]

    \addplot[
        color=cyan,
        mark=*,
    ]
    coordinates {
        (10,29.18)(25,32.14)(40,30.45)
    };

    \addplot[
        color=green,
        mark=*,
        dashed,
    ]     
    coordinates {
        (10,18.89)(25,17.41)(40,18.11)
    };

    \addplot[
        color=black,
        mark=*,
        dashdotted,
    ]
    coordinates {
        (10,0)(25,0)(40,0)
    };

    \addplot[
        color=red,
        mark=*,
        dash pattern=on 3pt off 2pt on 1pt off 2pt,
    ]
    coordinates {
        (10,4)(25,6)(40,7)
    };

    \end{axis}

\end{tikzpicture}
    \end{adjustbox}
    \caption{Smoothness}
    \label{plot:compare-smooth}
  \end{subfigure}


  \begin{subfigure}[t]{0.4\textwidth}
    \centering
    \begin{adjustbox}{width=\linewidth, height=0.25\textheight, keepaspectratio}
      \begin{tikzpicture}
    \begin{axis}[
        xlabel={{\LARGE \hspace{.7cm} \LARGE top-k \hspace{1cm} top-k-relaxed \hspace{1cm} top-k-affinity}}, 
        xlabel style={at={(axis description cs:0.5,0.07)}, anchor=north},
        ylabel={\LARGE Percentage (timeout)},
        ylabel style={at={(axis description cs:0.07,0.5)}, anchor=south},
        width=11cm,
        height=8cm,
        xmin=30, xmax=100,
        ymin=0, ymax=100,
        xtick={.},
        ytick={0,20,40,60,80,100},
        tick label style={font=\LARGE},
        legend pos=north west,
        legend entries={\large \textsc{Marabou ad-hoc encod - MNIST}, \large\textsc{Marabou layered encod - MNIST}, \large\textsc{\alphabeta{} layered encod - MNIST}, \large\textsc{\alphabeta{} layered encod - others}},
        ymajorgrids=true,
        grid style=dashed,
    ]

    \addplot[
        color=cyan,
        mark=*,
    ]
    coordinates {
        (40, 47.35)(65,46.55)(90,45.11)
    };

    \addplot[
        color=green,
        mark=*,
        dashed,
    ]
    coordinates {
        (40, 30.22)(65,31.15)(90,29.11)
    };

    \addplot[
        color=black,
        mark=*,
        dashdotted,
    ]
    coordinates {
        (40, 6.25)(65,7.78)(90,3.33)
    };

    \addplot[
        color=red,
        mark=*,
        dash pattern=on 3pt off 2pt on 1pt off 2pt,
    ]
    coordinates {
        (40,41.32)(65,39.34)(90,37.38)
    };

    \end{axis}
\end{tikzpicture}

    \end{adjustbox}
    \caption{\hspace{-1mm}Top-$k$ variants}
    \label{plot:compare-topk}
  \end{subfigure}%
  \hspace{0.5cm}
  \begin{subfigure}[t]{0.37\textwidth}
    \centering
    \begin{adjustbox}{width=\linewidth, height=0.15\textheight}
      \begin{tikzpicture}
    \begin{axis}[
        xlabel={\Large Number of benchmarks},
        xlabel style={at={(axis description cs:0.5,0.04)}, anchor=north},
        ylabel={\Large time(seconds)},
        ylabel style={at={(axis description cs:0.12,0.5)}, anchor=south},
        width=7cm,
        height=6cm,
        xmin=0, xmax=460,
        ymin=0, ymax=300,
        xtick={0, 50, 100, 200, 300, 400},
        ytick={0, 50, 100, 150, 200, 250, 300},
        legend style={
            at={(1,1.33)},
            draw=black, 
            fill=white, 
        },
        legend entries={
            \small \textsc{Marabou ad-hoc encod - MNIST},
            \small \textsc{Marabou layered encod - MNIST},
            \small \textsc{$\alpha \beta$-Crown layered encod - MNIST}
        },
        ymajorgrids=true,
        xmajorgrids=true,
        grid style=dashed,
    ]

    \addplot[color=cyan] table {figs/cactus/comparison/relaxed/marabou/data.txt};
    \addplot[color=green] table {figs/cactus/comparison/relaxed/marabou/data1.txt};
    \addplot[color=black] table {figs/cactus/comparison/relaxed/abcrown/data.txt};

    \end{axis}
\end{tikzpicture}
    \end{adjustbox}
    \caption{Cactus plots 
    }
    \label{fig:compare-cactus}
  \end{subfigure}
  \vspace{-3mm}
  \caption{
    (A)--(D) Comparison of the constraint-based solver \marabou{} with ad-hoc encoding and our layered-based encoding, alongside \alphabeta{} with our layered-based encoding. "\textsc{\alphabeta{} layered encod - others}" represents the average of percentage of timeout on other datasets. (E) the $x$-axis shows the number of benchmarks solved, ordered by increasing solving time, and the $y$-axis shows the time taken to solve them.
  }
  \label{fig:combined-comparison}
    \vspace{-6mm}
\end{figure}


{\bf Benchmarks:} 
We conducted experiments on four different datasets: MNIST \cite{deng2012mnist}, CIFAR-10~\cite{krizhevsky2009learning}, Traffic Sign Recognition (TSR)~\cite{postovan2023architecturing}, and IMAGENET~\cite{5206848}. All networks used in the experiments, along with their properties (\vnnlib{} files), were taken from \vnncomp{} 2021 to \vnncomp{} 2024. The provided \vnnlib{} files correspond to standard robustness properties. We have listed our benchmarks in~Table~\ref{tab:net_details} of Appendix~\ref{app:exp}. Our experiments covered a diverse set of benchmarks, ranging from fully connected networks to complex architectures, including convolutional layers, max-pooling layers, residual blocks, and \relu{} activations, including standard and adversarially trained models. {\em Additionally, the network sizes ranged from small architectures with $512$ ReLUs to large networks with $11.16$M ReLUs leading to the largest network having $138$M parameters. Each neural network, along with its corresponding property (VNNLIB) file, represents a single benchmark. In total, we evaluated 8,870 benchmarks in our experiments.} More details are provided in Appendix~\ref{app:benchmarks}.  We set the value of our encoding parameter $\smallparam$ to $1 \times 10^{-4}$.

{\bf Comparing with ad-hoc encoding (RQ1):} 
To compare our layer-based encoding with ad-hoc encodings, we directly encoded each property using the constraint-based solver \marabou{} via its Python interface. We then compared this against our layer-based encoding evaluated with \marabou{} (i.e., on the appended network with a simplified property) and with \alphabeta{}. 
Since \alphabeta{} does not provide an interface for directly expressing complex properties, we were unable to perform an ad-hoc encoding comparison for \alphabeta{}. However, our layer-based encoding enables \alphabeta{} to handle these properties.
As \marabou{} is a CPU-based solver~\cite{brix2024fifth}, we ran the experiments on an Intel(R) Xeon(R) Gold 6314U CPU @ 2.30\,GHz with 64\,GB RAM. We observed that \marabou{} ran out of memory on many CIFAR-10, GTSRB, and IMAGENET-1k benchmarks, so we restricted this comparison to MNIST. The blue, green and black lines in Figures~\ref{plot:compare-relaxed}--\ref{plot:compare-topk} summarize the comparison across all properties, and Figure~\ref{fig:compare-cactus} presents the corresponding cactus plot. 
  Our results show that \alphabeta{} with our layer-based encoding outperforms \marabou{} with both the ad-hoc and the layer-based encodings, over varying confidence values. This is a direct result of our layer-based approach being amenable for integration with \alphabeta{}, a portfolio tool that incorporates efficient techniques such as PGD attacks~\cite{dong2018boosting} and CROWN~\cite{zhang2018efficient}. However, we also see that our layer-based encoding on \marabou{} performs better than \marabou{} with the ad-hoc encoding. {\it This comparison demonstrates that our framework not only enables encoding of different properties, but also yields more efficient verification than direct encodings.}

{\bf Scalability evaluation of robustness variants (RQ2):}
We conducted experiments on various robustness properties, as described in Section~\ref{sec:examples}. The corresponding results are shown by the red dashed lines in Figures~\ref{plot:compare-relaxed}--\ref{plot:compare-topk} itself . These curves report results on CIFAR-10, GTSRB, and IMAGENET-1k, which are considered challenging benchmarks in \vnncomp{}. Accordingly, we used a GPU machine with a Tesla V100-SXM2-32GB, 9 vCPUs, and 32\,GB RAM for these experiments. The red curves represent the percentage of timeout instances across different thresholds and remain relatively low, indicating that a large fraction of benchmarks are successfully solved. More experiments comparing our approach across {\em varying thresholds} for different variants are in Appendix~\ref{app:exp:ablation}.

\section{Conclusion}
\label{sec:conclusion}
\vspace*{-2mm}

In this paper, we considered different variants of robustness in neural networks that incorporate \softmax{}-based confidence values. 
We introduced a grammar of post-conditions that captures all these variants and provided a unifying framework to transform them into a gadget that can be appended to an existing neural network, thereby simplifying arbitrary properties. This enables the use of verifiers such as \alphabeta{}. 
Our experiments demonstrate that our approach is more efficient than directly encoding the properties as constraints into state-of-the-art constraint-based solvers.


%

%

%

%


\bibliographystyle{unsrt}
\bibliography{biblio}







\section*{Appendix}

\appendix

\section{Softmax Approximation}  
\label{sec:analysis_compare}

In Figure~\ref{fig:lb-ub-approx}, we illustrate the estimation of the lower bound, by plotting confidence vs the gap between the top two outputs.
The thick curve plots the user defined thresholds, and the dashed curve plots the lower bound approximation.
For a target confidence threshold $\thresh$ on the Y-axis, a horizontal line to the user defined threshold curve finds the $\delta$, while the vertical line from the intersection crosses the lower bound curve at $\thresh_{lb}$; the lower bound for any counterexample.

\begin{figure}[t]
  \centering
\begin{tikzpicture}[domain=0:4,scale=1.4]
  \draw[very thin,color=gray] (-0.1,0) grid (4,1);

  \draw[->] (-0.2,0) -- (4.3,0) node[right] {$Gap$};
  \draw[->] (0,-.2) -- (0,1.2) node[above] {$Confidence$};

  \draw[-,thick] (0,.8) node[left] {$\thresh$} -- (1.4,0.8) ;
  \draw[-,thick] (1.4,.8)  -- (1.4,0) node[below] {$\delta$};
  \draw[-,thick] (0,.31) node[left] {$\thresh_{lb}$} -- (1.4,0.31) ;

  \draw[color=blue,thick]   plot (\x,{1/(1+exp(-\x))})    node[right,yshift=1.5mm] {User defined threshold};
  \draw[color=orange,dashed] plot (\x,{1/(1+9*exp(-\x))}) node[right,yshift=-1.5mm] {Lower bound};
\end{tikzpicture}

  \caption{Behavior of lower bound $\thresh_{lb}$ and the user defined threshold $\thresh \geq 50$ approximations of $softmaxC$.}
  \label{fig:lb-ub-approx}
\vspace*{-.5cm}
\end{figure}






We also present a theoretical comparison between our \softmax{} approximation and the approximation introduced in~\cite{athavale2024verifying}. Below is the definition of the \softmax{} function. We omit the factor of 100 from the equation, as it was used solely to convert probabilities into percentages:  

\begin{align*}
  \text{Softmax}(y_1,\dots,y_m, c)  &= \frac{e^{y_c}}{\sum_{i=1}^{m} e^{y_i}}.
\end{align*}


The paper~\cite{athavale2024verifying} introduced lower and upper bounds for the softmax function as follows. Let $c$ be the index of the maximum value, i.e., $c = \arg\max_{i=1}^{m} y_i$:  

\begin{align}
  \text{Sig}(y_c - \max_{i=1,i\neq c}^{m} (y_i) - \log(m-1))  \leq \text{Softmax}(y_1,\dots,y_m, c) \leq \text{Sig}(y_c - \max_{i=1,i\neq c}^{m} (y_i))
\end{align}

where $\text{Sig}(x) = \frac{1}{1+e^{-x}}$ is the sigmoid function, and $\max$ denotes the usual maximum operation.  

For analysis, consider the lower bound:  
\begin{align*}
  lb & = \text{Sig}(y_c - \max_{i=1,i\neq c}^{m} (y_i) - \log(m-1)) \\
     & = \frac{1}{1+e^{-y_c + \max_{i=1,i\neq c}^{m} (y_i) + \log(m-1)}} \\
     & = \frac{1}{1+e^{-y_c + \max_{i=1,i\neq c}^{m} (y_i)} e^{\log(m-1)}} \\
     & = \frac{1}{1+(m-1)e^{-y_c + \max_{i=1,i\neq c}^{m} (y_i)}} \\
     & = \frac{1}{1+(m-1)e^{-(y_{\max} - y_{\text{smax}})}}.
\end{align*}

Here, we define  
\begin{align*}
  y_{\max} &= \max_{i=1}^{m} y_i, \\
  y_{\text{smax}} &= \max_{i=1, i\neq \max}^{m} y_i.
\end{align*}
Intuitively, $y_{\max}$ and $y_{\text{smax}}$ represent the maximum and second maximum values of $y_i$, respectively.

We define the lower bound from Section~\ref{sec:examples} as follow:   
\begin{align*}
  lb' = \frac{1}{1+e^{-\delta}},  
\end{align*}
where $\delta = -\log\left(\frac{1}{th} - 1\right)$ and $\thresh$ is a user-defined confidence threshold.  

The key difference between our analysis and the one presented in~\cite{athavale2024verifying} is that our approach incorporates a user-defined confidence threshold, whereas the previous approximation is based solely on the input to softmax variables $y_i$.  

Now, let us determine when our lower bound approximation is tighter than the one from~\cite{athavale2024verifying}:  

\begin{align}
  \label{eq:lower:compare}
  lb' & > lb \notag \\
  \delta & > y_{\max} - y_{\text{smax}} \notag \\
  -\log\left(\frac{1}{th} - 1\right) & > y_{\max} - y_{\text{smax}} \notag \\
  \frac{1}{th} - 1 & < e^{-y_{\max} + y_{\text{smax}}} \notag \\
  \thresh & > \frac{1}{1+e^{-y_{\max} + y_{\text{smax}}}}. 
\end{align}

Eq~\ref{eq:lower:compare} establishes the condition under which our lower bound for the \softmax{} function is tighter. Applying a similar analysis for the upper bound, we find that  
\begin{align}
  \label{eq:upper:compare}
  \thresh & < \frac{1}{1+e^{-y_{\max} + y_{\text{smax}}}}.  
\end{align}  

Thus, depending on whether we are bounding the upper or lower limit, we can appropriately select $\thresh$ to achieve a tighter approximation.  

\section{Top-k Robustness}
\label{app:topk}
\begin{figure}[t]
    \centering
    \includegraphics[scale=0.3]{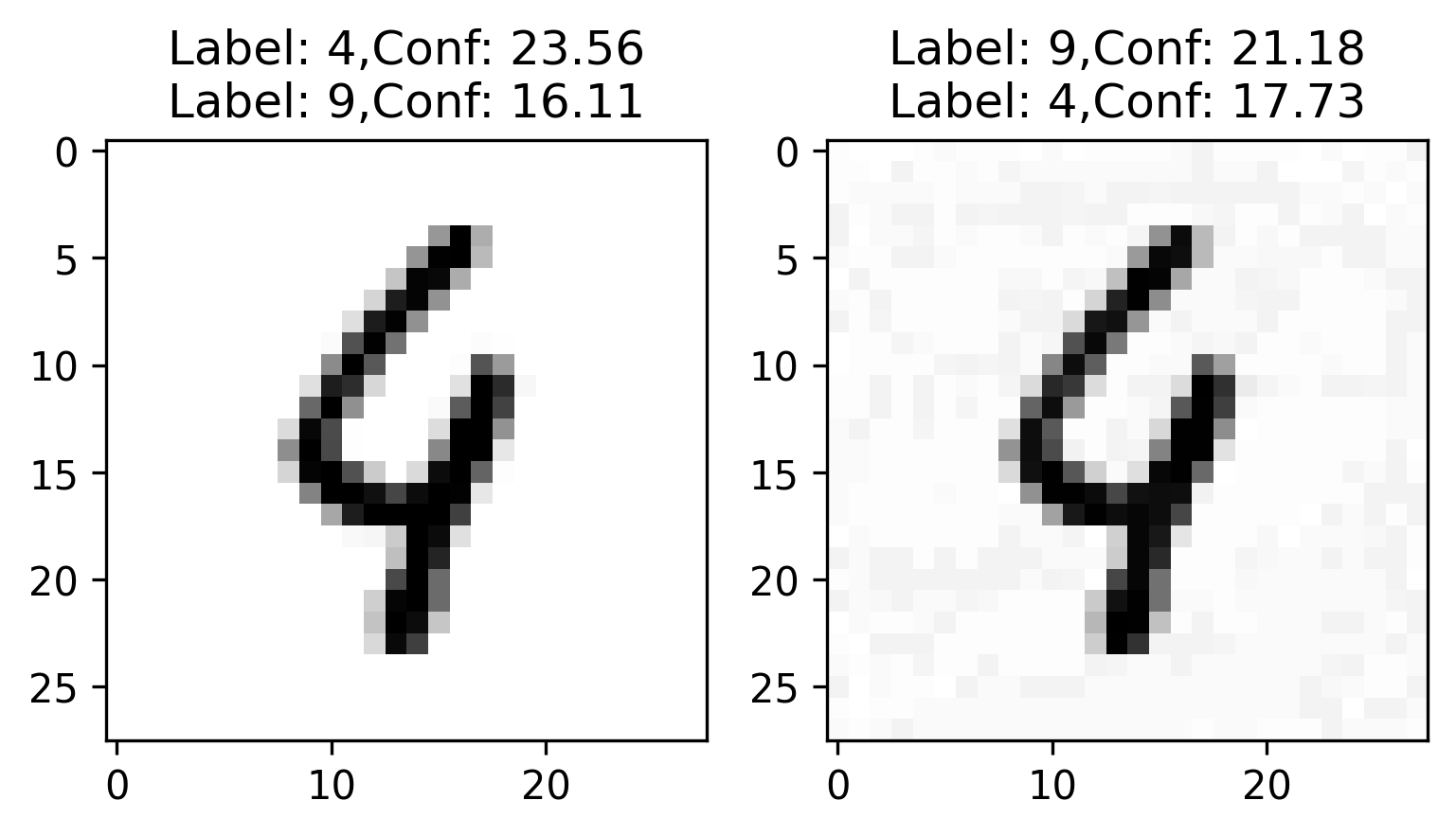}

    \caption{
      \textbf{Top-k and Top-k-relaxed}: The image on the left, labeled as $4$, is taken from the MNIST dataset and correctly classified with a confidence of $23.56\%$ by the neural network \texttt{mnist-net-256x6.onnx}. The tool~\alphabeta{} finds a counterexample (right), misclassified as label $9$, within an input perturbation of $0.05$. The counterexample has a confidence of $21.18\%$. The top-$k$ property allows misclassification within the top-$k$ predictions of the original image. For $k=2$, the original image has top-2 predictions: labels $4$ and $9$ (corresponding to the highest and second-highest confidence scores). This example is top-2 as well as top-2-relaxed robust but does not satisfy standard robustness or strong robustness criteria.
    }

	  \label{fig:topk}
  \end{figure}



Till now, we have considered robustness variants involving confidence, now we show that even other variants can easily be captured by our framework. We consider the notion of top-$k$ robustness introduced in~\cite{NEURIPS2021_8df6a659}. Suppose a function $N$ computes the logits values at the neural network's output layer, where  $N_i(x)$ represents the value at the $i^{th}$ dimension. Let $N(x,k)$ denote the $k^{th}$ highest value of the logits. The function $N^k(x) = \{i \mid N_i(x) \geq N(x,k)\}$ intuitively represents the set of classes with the top $k$ highest values of the logits. Top-$k$ robustness is then defined as: 
\begin{align*}
    \forall x' \; dist(x,x') \leq \epsilon \implies N^k(x) = N^k(x')
\end{align*} 

See Figure~\ref{fig:topk} for an example. Its negation can be expressed as:
\begin{align*}
    \exists x' \; dist(x,x') \leq \epsilon \land N^k(x) \neq N^k(x')
\end{align*}
Letting $y_1,...,y_m = N(x')$, we can capture this, i.e., $N^k(x) \neq N^k(x')$ constraint using
\begin{align}
   \label{app:eq:topk}
  \bigvee_{i=0}^{m}\bigvee_{ j \in N^k(x), j \neq i}^m y_i > y_j
\end{align}
which indeed is in our desired format for disjunction in the grammar, with the total number of disjunctive clauses is $(m-k) \cdot k$.

\begin{theorem}
  \label{app:th:topk}
  Eq~\eqref{app:eq:topk} $\Longleftrightarrow$ $N^k(x) \neq N^k(x')$
\end{theorem}

\begin{proof}
  ($\Longrightarrow$) If Eq~\ref{app:eq:topk} holds, for a fixed $k$ there exist indices $i$ and $j$ with $i \notin N^k(x)$ and $j \in N^k(x)$ such that $y_i > y_j$. By definition of $N^k(x)$, $i \in N^k(x')$, but $i \notin N^k(x)$ this means $N^k(x') \neq N^k(x)$.

  ($\Longleftarrow$) If for a fixed $k$, $N^k(x) \neq N^k(x')$ holds and $|N^k(x)| = |N^k(x')|$ then there must exist an index $i$ such that $i \in N^k(x')$ but $i \notin N^k(x)$.  Consequently, there also exists an index $j \in N^k(x)$ such that $j \notin N^k(x')$. This implies that $j$ is replaced by $i$ in $N^k(x')$, which further implies that $y_i > y_j$. Which is Eq~\ref{app:eq:topk}.
\end{proof}

\subsection{Top-k Relaxed Robustness:} 
Further, in the same paper~\cite{NEURIPS2021_8df6a659}, a relaxed variant of top-$K$ robustness was also defined as
\begin{align*}
    \forall x' \; dist(x, x') \leq \epsilon \implies \exists k \leq K : N^k(x) = N^k(x')
\end{align*}

The negation of the above post-condition is: 

\begin{align*}
    \bigwedge_{k=1}^{K} N^k(x) \neq N^k(x')
\end{align*}

Using the above, the full negation of the post-condition is expressed as:
\begin{align}
    \label{app:eq:topk:rel}
  \bigwedge_{k=1}^{K} \left( \bigvee_{i=1, j \in N^k(x), j \neq i}^{m} y_i > y_j \right)
\end{align}

The above constraints are in Conjunctive Normal Form (CNF) and hence expressible in our grammar and encodable using the procedure that we explain in Section~\ref{sec:framework}. 

\begin{theorem}
  \label{app:th:topk:rel}
  Eq~\ref{app:eq:topk:rel} $\Longleftrightarrow$ $\bigwedge_{k=1}^{K} N^k(x) \neq N^k(x')$
\end{theorem}

\begin{proof}
  ($\Longrightarrow$) If Eq~\ref{app:eq:topk:rel} holds, then for every $k \in K$ there exist indices $i$ and $j$ with $i \notin N^k(x)$ and $j \in N^k(x)$ such that $y_i > y_j$. By definition of $N^k(x)$, $i \in N^k(x')$, but $i \notin N^k(x)$ this means $N^k(x') \neq N^k(x)$ for all $k \in K$.

  ($\Longleftarrow$) If $\bigwedge_{k=1}^{K} N^k(x) \neq N^k(x')$ holds and $|N^k(x)| = |N^k(x')|$ then for each $k$ there must exist an index $i$ such that $i \in N^k(x')$ but $i \notin N^k(x)$.  Consequently, there also exists an index $j \in N^k(x)$ such that $j \notin N^k(x')$. This implies that $j$ is replaced by $i$ in $N^k(x')$, which further implies that $y_i > y_j$. Which is Eq~\ref{app:eq:topk:rel}.
\end{proof}

Theorem~\ref{app:eq:topk:rel} establishes the equivalence between the encoding constraints in Eq~\ref{app:eq:topk:rel} and the negation of the post-condition $\bigwedge_{k=1}^{K} N^k(x) \neq N^k(x')$ for relaxed Top-$k$ robustness. This equivalence confirms the correctness of our encoding.

\subsection{Top-k-affinity Robustness:}
\label{app:topk-aff}
The paper~\cite{NEURIPS2021_8df6a659} also introduced the concept of affinity robustness. The intuition behind this robustness is to incorporate expert knowledge into robustness evaluation. Experts provide sets of similar categories, and the network is considered robust as long as misclassifications remain within these predefined sets. The key idea is to allow misclassification within similar categories but not across completely different ones. 

For instance, if an input image of a pine tree is misclassified as a palm tree, this might be acceptable. However, it should not be misclassified as a mammal or another unrelated category. Similarly, a tiger could be misclassified as a leopard but not as an elephant. Suppose experts define these sets as $\mathbb{S}$, where each set $S \in \mathbb{S}$ represents a collection of classes within which misclassification is allowed.

\begin{figure}[t]
   \centering
    \begin{tabular}{cc}
      \scriptsize \texttt{label: 7} & \scriptsize \texttt{label: 9} \\
      \scriptsize \texttt{label: 9} & \scriptsize \texttt{label: 7} \\
      \includegraphics[scale=0.1]{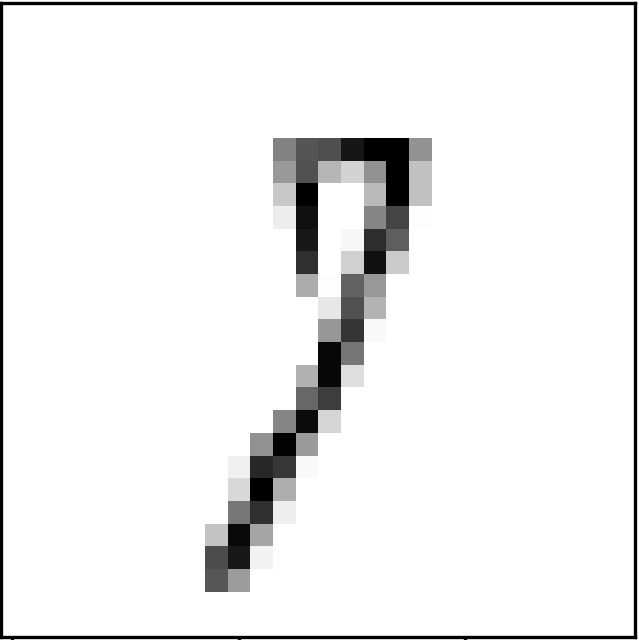} & \includegraphics[scale=0.1]{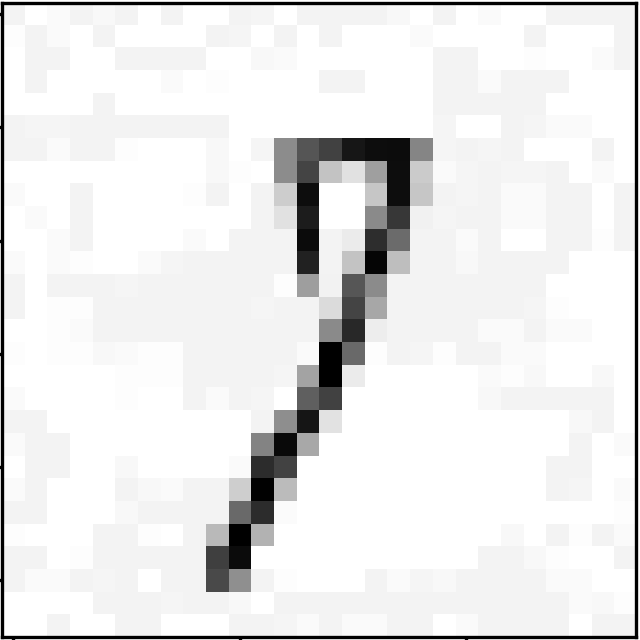} 
    \end{tabular}

  \caption{
    \textbf{Top-k-affinity Robustness}: The image on the left, labeled as $7$, is taken from the MNIST dataset and correctly classified with a confidence of $23.18\%$ by the neural network \texttt{mnist-net-256x6.onnx}. A state-of-the-art verification tool, \alphabeta{}, finds a counterexample (right), misclassified as label $9$, within an input perturbation of $0.05$ (consistent with the perturbation used in \vnncomp{}). The counterexample has a confidence of $21.34\%$.  
    In affinity robustness, the user provides prior knowledge about acceptable misclassifications. Suppose the user specifies the following affinity groups:  
    $\{\{1,7,9\}, \{2,7\}, \{0,8\}, \{4,8\}, \{3,5\}, \{6\}\}$.  
    This means, a class-$7$ image may be misclassified as $1$, $2$, or $9$. A class-$0$ image may only be misclassified as $8$. And a class-$6$ image must not be misclassified at all.  
    According to this specified misclassification knowledge, the above example is affinity robust as well as top-$k$ relaxed robust. However, the images in Figure~\ref{fig:topk} are not affinity robust under the same misclassification constraints because $4$ is not allowed to misclassified in class $9$. Additionally, the above image does not satisfy standard or strong robustness criteria.  
}
  \label{fig:affinity}
\end{figure}

The definition of affinity robustness is given as:

\begin{align}
    \forall x' \; & dist(x, x') \leq \epsilon \implies  \notag \\ 
    & \exists k \leq K \; S \in \mathbb{S} : N^k(x) = N^k(x') \land N^k(x) \subseteq S
    \label{eq:aff_r}
\end{align}

The negation of the post-condition can be written as follows:

\begin{align}
    \bigwedge_{k=1}^{K} \; \bigwedge_{S \in \mathbb{S}} \left( N^k(x) \neq N^k(x') \lor N^k(x) \nsubseteq S \right).
    \label{eq:aff_r_n}
\end{align}

For a given image $x$, the condition $N^k(x) \nsubseteq S$ can be determined beforehand. For all paired $\langle k, S \rangle$, if $N^k(x) \nsubseteq S$ is satisfied, the entire constraint is satisfied due to the disjunction. Therefore, we only need to construct constraints for those pairs $\langle k, S \rangle$ where $N^k(x) \subseteq S$ holds. Let $K'$ and $S'$ denote the sets of pairs $\langle k, S \rangle$ for which $N^k(x) \subseteq S$ is satisfied. The new constraints for the post-condition can be written as follows:

\begin{align*}
    \bigwedge_{\langle k, S \rangle \in (K', S')} N^k(x) \neq N^k(x').
\end{align*}

Using equation above, this can be rewritten as:

\begin{align}
    \label{app:eq:topk:aff}
    \bigwedge_{\langle k, S \rangle \in (K', S')} \left( \bigvee_{i=1, j \in N^k(x), j \neq i}^{m} y_i \geq y_j \right)
\end{align}



The above constraints are expressed in conjunctive normal form (CNF) and can be encoded into a neural network using the procedure described in Section~\ref{sec:framework}.

\begin{theorem}
  \label{app:th:topk:aff}
  Eq~\ref{app:eq:topk:aff} $\Longleftrightarrow$ $\bigwedge_{\langle k, S \rangle \in (K', S')} N^k(x) \neq N^k(x')$
\end{theorem}

\begin{proof}
  ($\Longrightarrow$) If Eq~\ref{app:eq:topk:aff} holds, then for every $k,S \in (K',S')$ there exist indices $i$ and $j$ with $i \notin N^k(x)$ and $j \in N^k(x)$ such that $y_i > y_j$. By definition of $N^k(x)$, $i \in N^k(x')$, but $i \notin N^k(x)$ this means $N^k(x') \neq N^k(x)$ for all $k,S \in (K',S')$.

  ($\Longleftarrow$) If $\bigwedge_{\langle k, S \rangle \in (K', S')} N^k(x) \neq N^k(x')$ holds and $|N^k(x)| = |N^k(x')|$ then for each $k,S \in (K',S')$ there must exist an index $i$ such that $i \in N^k(x')$ but $i \notin N^k(x)$.  Consequently, there also exists an index $j \in N^k(x)$ such that $j \notin N^k(x')$. This implies that $j$ is replaced by $i$ in $N^k(x')$, which further implies that $y_i > y_j$ for each $k,S \in (K',S')$. Which is Eq~\ref{app:eq:topk:aff}.
\end{proof}

\section{Proof from Section~\ref{sec:examples}}
\label{sec:examples:proofs}

\begin{proof}[of Claim~\ref{claim:greater_ineq} part 1]
  The following proof is similar to the proof of part 2 of Claim~\ref{claim:less_ineq}. Let LHS of the claim be  true. Since $y_{t'} = \max_{i=1, i \neq t}^{m} (y_i)$, we can derive 
  $
  \Conf(\bar{y},t) \geq \frac{100e^{y_t}}{e^{y_{t}}+(m-1)e^{y_{t'}}},
  $
  replacing $e^i$ by $e^{t'}$ for all $i,i\neq t$. 
  After scaling the fraction in the RHS by $e^{y_t}$, we obtain
  $
  \Conf(\bar{y},t) \geq \frac{100}{1+(m-1)e^{y_{t'} - y_t}}.
  $
  Since $y_t \geq y_{t'} + \delta$, we obtain
  $
  \Conf(\bar{y},t) \geq \frac{100}{1+(m-1)e^{-\delta}}.
  $
  Since $\delta = -\ln (\frac{1}{m-1} (\frac{100}{b} - 1 ) )$, 
  $
  \Conf(\bar{y},t) > b.
  $  
\end{proof}


\begin{proof}[of Claim~\ref{claim:greater_ineq} part 2]
  The following proof is similar to the proof of part 1 of Claim~\ref{claim:less_ineq}. Assume LHS of the claim holds, i.e., $y_t \leq y_{t'} + \delta$. By removing all exponential terms from the denominator except for $e^{y_t}$ and $e^{y_{t'}}$, we obtain 
  $
  \Conf(\bar{y},t) \leq \frac{100e^{y_t}}{e^{y_{t}}+e^{y_{t'}}}.
  $
  After scaling the right-hand side by $\frac{1}{e^{y_t}}$, we get 
  $
  \Conf(\bar{y},t) \leq \frac{100}{1+e^{-(y_t - y_{t'})}}.
  $ 
  Since $\delta \geq y_t- y_{t'}$, we have
  $
  \Conf(\bar{y},t) \leq \frac{100}{1+e^{-\delta}}.
  $ 
\end{proof}

\begin{proof}[of Theorem~\ref{th:grammar:soundness}]
  We prove soundness by showing that whenever $PC^\#$ holds, the original post-condition $PC$ also holds. The proof proceeds by induction on the length $n$ of $PC$. If $PC$ does not contain any confidence constraint $CC$, then $PC$ and $PC^\#$ are identical. We therefore focus on cases where $PC$ contains confidence constraints.
  \begin{itemize}
    \item \textbf{Base Case ($n = 1$):} In this case, $PC = CC$, which implies $PC^\# = CC^\#$. By Claims~\ref{claim:less_ineq} and~\ref{claim:greater_ineq}, we have $CC^\# \implies CC$, and thus $PC^\# \implies PC$.
    \item \textbf{Inductive Step:} Assume the theorem holds for all post-conditions of length $n$. Let $PC_n$ be a post-condition of length $n$, and let $PC^\#_n$ be its corresponding abstraction. We now consider $PC_{n+1}$ and its abstraction $PC^\#_{n+1}$, which form a post-condition of length $n+1$. The structure of $PC_{n+1}$ must be one of the following:
    \begin{enumerate}
      \item $PC_{n+1} = PC_n \land CC$.  
      Then $PC^\#_{n+1} = PC^\#_n \land CC^\#$.  
      Since $CC^\# \implies CC$ (by Claims~\ref{claim:less_ineq} and~\ref{claim:greater_ineq}) and $PC^\#_n \implies PC_n$ (by the inductive hypothesis), we obtain $PC^\#_n \land CC^\# \implies PC_n \land CC$. 
      \item $PC_{n+1} = PC_n \lor CC$.  
      Then $PC^\#_{n+1} = PC^\#_n \lor CC^\#$.  
      Since $CC^\# \implies CC$ and $PC^\#_n \implies PC_n$, it follows that $PC^\#_n \lor CC^\# \implies PC_n \lor CC$.  
    \end{enumerate}
  \end{itemize}
\end{proof}

\section{Section~\ref{sec:framework}: Proofs of Correctness}
\label{sec:framework-proofs}



  \begin{proof}[Proof for Lemma \ref{thm:encoding_conj}]
($\implies$)
    Since $y = \sum_{i=0}^{n} \relu{}(LE_i)$,
    if $y \leq 0$ then
    $\forall i$, $\relu{}(LE_i) = 0$, because \relu{}'s output is always $\geq 0$.
    Therefore, $\forall i, LE_i \leq 0 $ by the definition of \relu{}. Therefore,
    $\Land_{i=0}^{n} LE_i \leq 0$ holds.
 ($\Longleftarrow$)
    Since $\Land_{i=0}^{n} LE_i \leq 0$, we conclude $\Land_{i=0}^{n} \relu{}(LE_i) = 0$.
    Therefore, $\sum_{i=0}^{n} \relu{}(LE_i) = 0$.
    Since $y = \sum_{i=0}^{n} \relu{}(LE_i)$, $y \leq 0$. \qed
  \end{proof}

\begin{proof} [Proof for Lemma \ref{thm:ub3}]

    The following is the proof of the four parts.
    \begin{enumerate}
    \item
      We assume for each $i$, $\smallparam \leq V(\lor,Q_i,\smallparam)$.
      After a rewrite, $(1+1/k)\smallparam- V(\lor,Q_i,\smallparam) \leq \smallparam/k$.
      After applying definition of $b$ and $flip$,
      $flip(b(\land,k,\smallparam), V(\lor,Q_i,\smallparam)) \leq \smallparam/k$ for each $i$.
     Due to the definition of $V$, we conclude $V(\land,Q,\smallparam) \leq \smallparam $.
   \item 
      We assume for some $i$, $ V(\land,Q_i,\smallparam) \leq \smallparam$.
      After a rewrite, $ \smallparam \leq 2\smallparam- V(\lor,Q_i,\smallparam)$.
      After applying definition of $b$ and $flip$,
      $\smallparam \leq flip(b(\lor,k,\smallparam), V(\land,Q_i,\smallparam))$ for some $i$.
      Due to the definition of $V$, we conclude $\smallparam \leq V(\land,Q,\smallparam)$.
  \end{enumerate}
\end{proof}

\begin{proof}[of Lemma \ref{lemm:full}]
      We prove this lemma by induction.
    \begin{itemize}
      \item \textbf{Base case:}  
      \begin{enumerate}
        \item If $Q = LE \leq 0$, then by definition, $V(\land, Q, \smallparam) = LE + \smallparam$. Given that $V(\land, Q, \smallparam) \leq \smallparam$, we derive: $LE + \smallparam \leq \smallparam \lequiv LE \leq 0$.
        \item If $Q = LE \leq 0$, then by definition, $V(\lor, Q, \smallparam) = -LE + \smallparam$. Given that $V(\lor, Q, \smallparam) \geq \smallparam$, we derive: $-LE + \smallparam \geq \smallparam \lequiv  LE \leq 0$.
      \end{enumerate}
    \item \textbf{Inductive step:} Assume Lemma~\ref{lemm:full} holds for all formulae of depth $d$, denoted as $Q^d$. Consider a formula of depth $d+1$, given by: $Q^{d+1} = Q^{d}_1 \dagger \dots \dagger Q^{d}_k$.
      \begin{enumerate}
        \setcounter{enumi}{2}
      \item  If $\dagger = \land$, $Q_i^{d}$s are true.
        Due to the induction hypothesis, $\forall i, \;\smallparam \leq V(\lor, Q^d_i, \smallparam)$ .
        By the part 1 of Lemma~\ref{thm:ub3}, we have: $\forall i, \;\smallparam \leq V(\lor, Q^d_i, \smallparam) \implies V(\land, Q^{d+1}, \smallparam) \leq \smallparam $. Therefore, $V(\land, Q^{d+1}, \smallparam) \leq \smallparam$ holds.
      \item  If $\dagger = \lor$, $Q_i^{d}$ is true for some $i$.
        Due to the induction hypothesis, $\exists i, \; V(\land, Q^d_i, \smallparam) \leq \smallparam $ . By the part 2 of Lemma~\ref{thm:ub3}, we have: $\exists i, \; V(\land, Q^d_i, \smallparam) \leq \smallparam  \implies \smallparam \leq V(\lor, Q^{d+1}, \smallparam)   $. Therefore, $\smallparam \leq V(\lor, Q^{d+1}, \smallparam)$ holds.
      \end{enumerate}  
    \end{itemize}
  \end{proof}

  \begin{proof}[of Theorem~\ref{thm:query}]
  The following is the proof of the first part.
  Let us consider the successful query $\langle N', P, y < \smallparam \rangle$ to the solver.
    We know $\lnot ( y < \smallparam)$ is infeasible.
    Therefore, $\smallparam \leq V(\lor, Z, \smallparam)$ is infeasible.
    Due to the contrapositive of the fourth part of
    Lemma~\ref{lemm:full},
    we conclude that $Z$ cannot be satisfied.
    Since $Z$ is the negation normal form of $\lnot Q$, $Q$ always holds. 
    Therefore, query $\langle N, P, Q \rangle$ holds.
    Similarly, we can prove the second part using the last part of Lemma~\ref{lemm:full}. 
  \end{proof}

\begin{proof}[of Theorem~\ref{thm:final}]
  Let $ Q''$ denote an intermediate post-condition obtained after applying confidence approximations, but prior to encoding.
  From Theorems~\ref{th:grammar:soundness}, we have:
$  \langle N, P,  Q'' \rangle \implies \langle N, P,  Q \rangle$.
  Furthermore, from Theorem~\ref{thm:query}, we have:
$  \langle N', P,  Q' \rangle \implies \langle N, P,  Q'' \rangle.$
and hence together, we conclude
$  \langle N', P,  Q' \rangle \implies \langle N, P,  Q \rangle.$
\end{proof}

 \paragraph{Bound on the error in counterexamples.}
  If the query $\langle N', P, y < \smallparam \rangle$ fails, the verifier
  generates a counterexample input $x$ to $N'$ that violates $y < \smallparam$. 
  We need to prove the following theorems~\ref{thm:ub4}, which is similar to
  Lemma~\ref{thm:ub3} but the direction bounds flow reverse.
\begin{lemma}For $\smallparam > 0$ and $\beta > 0$, the following holds for $Q$.
    \label{thm:ub4}
  \begin{enumerate}
\item  
  $V(\land,Q,\smallparam) \leq \beta \limplies \forall i.\; (1+1/k)\smallparam-\beta \leq V(\lor,Q_i,\smallparam)$.
\item   
  $ \beta \leq V(\lor,Q,\smallparam) \limplies \exists i.\; V({\land},Q_i,\smallparam) \leq 2\smallparam-\beta/k$.
  \end{enumerate}
\end{lemma}

\begin{proof}
  \begin{enumerate}
    \item
      If $V(\land,Q,\smallparam) \leq \beta $, we conclude $flip(b(\land,k,\smallparam), V(\lor,Q_i,\smallparam)) \leq \beta$.
      After applying definition of $b$,
      $flip((1+1/k)\smallparam, V(\lor,Q_i,\smallparam)) \leq \beta$ for each $i$.
      After expanding definition of $flip$, $(1+1/k)\smallparam- V(\lor,Q_i,\smallparam) \leq \beta$.
      After simplification, $(1+1/k)\smallparam -\beta \leq V(\lor,Q_i,\smallparam)$.
   \item 
      If $\beta \leq V(\lor,Q,\smallparam) $, we conclude $\beta/k \leq flip(b(\land,k,\smallparam), V(\land,Q_i,\smallparam))$ for some $i$.
      After applying definition of $b$,
      $\beta/k \leq flip(2\smallparam, V(\land,Q_i,\smallparam)) $.
      After expanding definition of $flip$,
      $\beta/k \leq 2\smallparam - V(\land,Q_i,\smallparam) $.
      After simplification, $V(\land,Q_i,\smallparam) \leq 2\smallparam -\beta/k$.
    \end{enumerate}
\end{proof}
  Since the neural network translation is approximate, $x$ may not violate $\lnot Q$ in $N$.
  Moreover, the following Theorem 
  uses the above theorem to bound the error.
  Let $Q[\smallparam]$ be a formula obtained by replacing  each $LE \leq 0$ by $LE \leq \smallparam$ in $Q$.
  \begin{theorem}
    \label{thm:sat}
  If $ \smallparam \leq V(\lor, Q, \smallparam)$ is satisfiable and $Q$ is DNF (disjunctive normal form), we can conclude that $~Q[2\smallparam]$ is satisfiable.    
\end{theorem}

  \begin{proof} 
  Since $Q$ is DNF, $Q = Q_1 \lor ... \lor Q_k $.
  Due to part four of Lemma~\ref{thm:ub3},
  there is $V({\land},Q_i,\smallparam) \leq 2\smallparam-\smallparam/k = (2-1/k)\smallparam$.
  Let $Q_i = LE_1 \leq 0 \land ... \land LE_l \leq 0$. 
  Due to part three of Lemma~\ref{thm:ub3} for each $LE_i$,
  $(1+1/l)\smallparam-(2-1/k)\smallparam \leq V(\lor,LE_i \leq 0,\smallparam)$.
  After simplification,
  $(-1+1/k+1/l)\smallparam\leq V(\lor,LE_i \leq 0,\smallparam)$.
  Therefore, $(-1+1/k+1/l)\smallparam \leq -LE_i+ \smallparam$.
  Therefore, $LE_i \leq (2-1/k-1/l)\smallparam \leq 2\smallparam$.
\end{proof}

The above theorem indicates that the satisfying assignment may violate $Q$ by the margin of $2\smallparam$. We must choose $\smallparam$ as small as possible
but not smaller than the minimum precision of the underlying solver, which will lead to the numerical instability of the solver. The above theorem requires $Q$ to be in DNF. However, we can prove a similar theorem if $Q$ is in CNF (conjunctive normal form).
Since all our properties are in CNF or DNF, we can use the above theorem to bound the error of the counterexample for the properties that we are interested in.

\section{Support of strict inequalities}
\label{sec:support-strict}

In this section, we present the version of theorem from section 4 that support strict inequalities.
The proof of theorems work in the similar lines as we have provided for the earlier theorems.

We can rewrite Lemma~\ref{thm:ub3} as follows to support strict inequalities. Let $<^* \in \{<,\leq\} $. Let us define $V$ for $<^*$ : 
$$
  V(\land, LE <^* 0,\smallparam) = LE+\smallparam \qquad 
  V(\lor, LE <^* 0,\smallparam) = -LE+\smallparam 
$$

\begin{lemma} For $\smallparam > 0$ and $\beta > 0$, the following holds for $Q$. 
    \label{thm:ub3-strict}
  \begin{enumerate}
\item  
  $\forall i.\;\smallparam <^*_i V(\lor,Q_i,\smallparam) \limplies V(\land,Q,\smallparam) <^* \smallparam $.\\
  If for some $i$, $<^*_i = <$, then $<^* = <$. Otherwise, $<^* = \leq$.   
\item   
  $\exists i.\; V({\land},Q_i,\smallparam) <^* \smallparam \limplies \smallparam <^* V(\lor,Q,\smallparam)$.
\item  
  $V(\land,Q,\smallparam) <^* \beta \limplies \forall i.\; (1+1/k)\smallparam-\beta <^* V(\lor,Q_i,\smallparam)$.
\item   
  $ \beta <^* V(\lor,Q,\smallparam) \limplies \exists i.\; V({\land},Q_i,\smallparam) <^* 2\smallparam-\beta/k$.
\end{enumerate}
\end{lemma}

Let us recursively define $<_{Q}^*$.
\begin{enumerate}
\item For $Q = LE <^* 0$, $<_Q^* = <^*$.
\item For conjunctive $Q$
$$<_{Q}^* =
\begin{cases}
  \leq & \forall i. <_{Q_i} = \leq \\
  < & Otherwise
\end{cases}
$$
\item For disjunctive $Q$
$$<_{Q}^* =
\begin{cases}
  < & \forall i. <_{Q_i} = < \\
  \leq & Otherwise
\end{cases}
$$
\end{enumerate}

\begin{lemma}
  \label{lemm:full-strict}
  For a given post condition $Q$ and an $\smallparam> 0$, the following holds
\end{lemma}
    \begin{enumerate}
  \item If $Q = LE \leq 0$ , $V(\land, Q, \smallparam) \leq \smallparam \lequiv Q$  
  \item If $Q = LE \leq 0$ , $V(\lor,  Q, \smallparam) \geq \smallparam \lequiv Q$
  \item {If $Q = LE > 0$ , $V(\land, Q, \smallparam) < \smallparam \lequiv Q$}
  \item {If $Q = LE > 0$ , $V(\lor,  Q, \smallparam) > \smallparam \lequiv Q$}
  \item If $Q$ is conjunctive, $Q \limplies V(\land, Q,\smallparam) <_{Q}^* \smallparam$
  \item If $Q$ is disjunctive, $Q \limplies \smallparam <_{Q}^* V(\lor, Q,\smallparam)$
  \end{enumerate}

  Let $\overline{<} = \leq$ and $\overline{\leq} = <$.
  
  \begin{theorem}
    \label{thm:query-strict}
    Let us consider a neural network $N$, pre-condition $P$, a post-condition $\lnot Q$,
    and $\smallparam > 0$.
    \begin{enumerate}
      \item If $Q$ is disjunctive, $\langle N', P, y \overline{<_{Q}^{*}} \smallparam \rangle \limplies \langle N, P, \lnot Q \rangle$
\item If $Q$ is conjunctive, $ \langle N', P,  \smallparam \overline{<_{Q}^{*}} y \rangle \limplies \langle N, P, \lnot Q \rangle$
    \end{enumerate}
  \end{theorem}

\section{Experiments}
\label{app:exp}
In this section, we show the detailed analysis of the Experiments in section~\ref{sec:experiments}.

\subsection{Benchmarks}
\label{app:benchmarks}
The network details are summarized in Table~\ref{tab:net_details}. In this section, we briefly describe each benchmark as follows: 

\begin{table}[t]
  \centering
  \caption{Networks details}
  \label{tab:net_details}
  \scalebox{0.8}{
    \begin{tabular}{|c|c|c|c|c|c|c|}
    \hline
    \textbf{Category} & \textbf{Network name} & \#layers & \#activation units & adv trained  \\
    \hline
    MNIST & mnist-net-256$\times$2.onnx  & 2-FC & 0.51K & No  \\
    \hline
     & mnist-net-256$\times$4.onnx  & 4-FC & 1.02K & No   \\
    \hline
     & mnist-net-256$\times$6.onnx  & 6-FC & 1.54K & No  \\
    \hline
    CIFAR-10 & cifar-base-kw.onnx  & 2-Conv, 2-FC & 3.17K & Yes \\
    \hline
     & cifar-deep-kw.onnx & 4-Conv, 2-FC & 6.77K & Yes \\
    \hline
     & cifar-wide-kw.onnx & 2-Conv, 2-FC & 6.24K & Yes \\
    \hline
     & cifar10-2-255.onnx  & 3-Conv, 2-FC & 49.15K & Yes  \\
    \hline
     & cifar10-8-255.onnx & 2-Conv, 2-FC & 16.39K & Yes  \\
    \hline
     & convBigRELU-PGD.onnx & 4-Conv, 3-FC & 62.46K & Yes  \\
    \hline
     & resnet-2b.onnx & 2-res-blocks (5-Conv, 2-FC) & 6.24K & Yes \\
    \hline
     & resnet-4b.onnx & 4-res-blocks (9-Conv, 2-FC) & 14.45K & Yes  \\
    \hline 
    GTSRB & net-1  & 2-QConv, 1-FC & - & No \\ 
    \hline
     & net-2  & 3-QConv, 3-BN, 2-Maxpool, 2-FC & - & No   \\ 
    \hline
     & net-3  & 3-QConv, , 3-BN, 3-Maxpool, 2-FC & - & No  \\ 
    \hline
    IMAGENET & vggnet-16  & 13-Conv, 5-Maxpool, 3-FC & 13.16M & No \\
    \hline

\end{tabular}
  }
\end{table}

\noindent{\bf MNIST:}  
We utilized benchmarks from \vnncomp{} 2022, which included a total of three networks and 30 \vnnlib{} files, resulting in 90 benchmarks. These benchmarks were constructed with $l_\infty$ input perturbations of $\epsilon \in \{0.03, 0.05\}$ in $15$ randomly choosen images from mnist dataset. All networks are fully connected with \relu{} activation functions, featuring between $0.51K$ and $1.54K$ ReLU activations.  

\noindent{\bf CIFAR-10:}  
We selected the first six networks, ranging from \texttt{cifar-base-kw.onnx} to \texttt{convBigRELU-PGD.onnx}, from \vnncomp{} 2022. These networks include convolutional layers, fully connected layers, and \relu{} activation functions. Additionally, the networks \texttt{resnet-2b} and \texttt{resnet-4b}, both residual networks~\cite{he2016deep} (ResNet), were sourced from \vnncomp{} 2021. These residual networks contain two and four residual blocks, respectively. All networks for this dataset were adversarially trained either using COLT~\cite{madry2018towards} or using the method described in~\cite{Balunovic2020Adversarial}. The dataset includes a total of 305 benchmarks, with $l_\infty$ input perturbations $\epsilon$ ranging from $\frac{1}{255}$ to $\frac{16}{255}$.

\noindent{\bf GTSRB:}  
These benchmarks are sourced from \vnncomp{} 2024 and are based on binary neural networks (BNNs) trained on the German Traffic Sign Recognition Benchmark (GTSRB) dataset~\cite{6033395}. This multi-class dataset comprises images of German road signs spanning 43 classes, presenting challenges for both humans and models due to factors such as perspective changes, shading, color degradation, and varying lighting conditions. The networks include QConv layers (which binarize the corresponding convolutional layers), Batch Normalization (BN), Max Pooling (MP), and Fully Connected (FC) layers. A total of 45 benchmarks are available for this dataset, created using $l_\infty$ input perturbations with parameter epsilon values of $\{1, 3, 5, 10, 15\}$. These networks do not use explicit activation functions but instead employ a \texttt{sign} operator, which converts the input to $1$, $-1$, or $0$ if the input is positive, negative, or zero, respectively.

\noindent{\bf IMAGENET-1K:}  
To analyze the effect on scalability when using appended networks, we used the VGGNET-16 architecture~\cite{simonyan2014very}, the only one from \vnncomp{} that runs on imagenet. The network consists of convolutional layers, \relu{} activation functions, and max pooling layers, with a total of $138M$ parameters and $13.16M$ \relu{} activations.  The properties were generated by applying perturbations on single input pixels to all 150528 input pixels using $l_\infty$ perturbations. The post condition was generated wrt target robustness, where we check for misclassification wrt a fixed target class.

\subsection{Ablation Study}
\label{app:exp:ablation}
In this section, we perform an ablation study by varying the confidence threshold. For example, increasing the value of \(\thresh\) should increase the number of safe cases under relaxed robustness. We conduct similar sanity-check analyses for the other properties as well. The results reported in this section are aggregated over all datasets. 

\begin{figure}[t]
  \centering
  \begin{subfigure}[t]{0.4\textwidth}
    \centering
    \begin{adjustbox}{width=\linewidth, height=0.4\textheight, keepaspectratio}
      \begin{tikzpicture}
    \begin{axis}[
        ylabel={\LARGE Percentage},
        ylabel style={at={(axis description cs:0.07,0.5)}, anchor=south},
        width=11cm,
        height=8cm,
        xmin=0, xmax=100,
        ymin=0, ymax=100,
        xtick={0, 60, 80, 90, 95},
        ytick={0,20,40,60,80,100},
        tick label style={font=\Large},
        yticklabel style={font=\LARGE},
        legend pos=north west,
        legend entries={\LARGE \textsc{unsafe}, \LARGE\textsc{safe}, \LARGE\textsc{timeout}},
        ymajorgrids=true,
        grid style=dashed,
    ]

    \addplot[
        color=cyan,
        mark=*,
    ]
    coordinates {
        (0,22.93)(60,21.18)(80,13.10)(90,11.14)(95,10.48)
    };

    \addplot[
        color=green,
        mark=*,
        dashed,
    ]
    coordinates {
        (0,63.10)(60,67.69)(80,81.44)(90,84.72)(95,85.37)
    };

    \addplot[
        color=black,
        mark=*,
        dashdotted,
    ]
    coordinates {
        (0,12.01)(60,11.14)(80,5.46)(90,4.15)(95,4.15)
    };

    \end{axis}

\end{tikzpicture}
    \end{adjustbox}
    \caption{Relaxed}
    \label{plot:relaxed1}
  \end{subfigure}%
  \qquad
  \begin{subfigure}[t]{0.4\textwidth}
    \centering
    \begin{adjustbox}{width=\linewidth, height=0.4\textheight, keepaspectratio}
      \begin{tikzpicture}
    \begin{axis}[
        ylabel={\LARGE Percentage},
        ylabel style={at={(axis description cs:0.07,0.5)}, anchor=south},
        width=11cm,
        height=8cm,
        xmin=0, xmax=42,
        ymin=0, ymax=100,
        xtick={0, 30, 35, 40},
        ytick={0,20,40,60,80,100},
        tick label style={font=\LARGE},
        legend pos=north west,
        legend entries={\LARGE \textsc{unsafe}, \LARGE\textsc{safe}, \LARGE\textsc{timeout}},
        ymajorgrids=true,
        grid style=dashed,
    ]

    \addplot[
        color=cyan,
        mark=*,
    ]
    coordinates {
        (0,15.70)(30,31.14)(35,34.43)(40,36.46)
    };

    \addplot[
        color=green,
        mark=*,
        dashed,
    ]
    coordinates {
        (0,71.65)(30,55.19)(35,53.16)(40,51.14)
    };

    \addplot[
        color=black,
        mark=*,
        dashdotted,
    ]
    coordinates {
        (0,12.66)(30,13.67)(35,12.41)(40,12.41)
    };

    \end{axis}

\end{tikzpicture}
    \end{adjustbox}
    \caption{Strong}
    \label{plot:strong1}
  \end{subfigure}%

  \begin{subfigure}[t]{0.4\textwidth}
    \centering
    \begin{adjustbox}{width=\linewidth, height=0.4\textheight, keepaspectratio}
      \begin{tikzpicture}
    \begin{axis}[
        ylabel={\LARGE Percentage},
        ylabel style={at={(axis description cs:0.07,0.5)}, anchor=south},
        width=11cm,
        height=8cm,
        xmin=5, xmax=45,
        ymin=0, ymax=100,
        xtick={5, 10, 25, 40},
        ytick={0,20,40,60,80,100},
        tick label style={font=\LARGE},
        legend style={at={(0.4,0.8)}, anchor=north east},  
        legend entries={\LARGE \textsc{unsafe}, \LARGE\textsc{safe}, \LARGE\textsc{timeout}},
        ymajorgrids=true,
        grid style=dashed,
    ]

    \addplot[
        color=cyan,
        mark=*,
    ]
    coordinates {
        (10,89.37)(25,83.54)(40,78.48)
    };

    \addplot[
        color=green,
        mark=*,
        dashed,
    ]
    coordinates {
        (10,10.13)(25,13.67)(40,15.70)
    };

    \addplot[
        color=black,
        mark=*,
        dashdotted,
    ]
    coordinates {
        (10,0.51)(25,2.78)(40,6.58)
    };

    \end{axis}

\end{tikzpicture}
    \end{adjustbox}
    \caption{Smoothness}
    \label{plot:smooth1}
  \end{subfigure}%
  \qquad
  \begin{subfigure}[t]{0.4\textwidth}
    \centering
    \begin{adjustbox}{width=\linewidth, height=0.4\textheight, keepaspectratio}
      \begin{tikzpicture}
    \begin{axis}[
        width=9.8cm, height=7cm,
        ybar, 
        bar width=8pt, 
        ymin=0, ymax=100, 
        symbolic x coords={standard, top-k, top-k-rel, top-k-aff}, 
        xtick=data,
        nodes near coords, 
        every node near coord/.append style={font=\normalsize, black}, 
        enlarge x limits=0.2, 
        ylabel={\LARGE Percentage},
        ylabel style={at={(axis description cs:0.08,0.5)}, anchor=south},
        legend style={at={(0.8,0.99)}, anchor=north}, 
        xticklabel style={rotate=0, anchor=center, font=\Large}, 
        yticklabel style={font=\LARGE},
        ymajorgrids=true, grid style=dashed, 
        x tick label style={yshift=-5pt}, 
    ]
    
    \addplot[color=yellow, fill=yellow, bar shift=-15pt] coordinates {
        (standard,15.7) (top-k,13.39) (top-k-rel,9.37) (top-k-aff,11.39) 
    };
    
    \addplot[color=cyan, fill=cyan, bar shift=0pt] coordinates {
        (standard,71.65) (top-k,54.49) (top-k-rel,58.48) (top-k-aff,58.99)
    };
    
    \addplot[color=green, fill=green, bar shift=15pt] coordinates {
        (standard,12.66) (top-k,32.12) (top-k-rel,32.15) (top-k-aff,29.62)
    };
    
    \legend{\LARGE \unsafe{}, \LARGE\safe{}, \LARGE\timeout{}}
    
    \end{axis}
\end{tikzpicture}
    \end{adjustbox}
    \caption{Top-k}
    \label{plot:bargraph}
  \end{subfigure}
  \vspace{-3mm}
  \caption{
  Figures~\ref{plot:relaxed1}, \ref{plot:strong1}, and \ref{plot:smooth1} show the confidence thresholds on the x-axis and the percentage of \safe{}, \unsafe{}, and \timeout{} instances on the y-axis. Figure~\ref{plot:bargraph} presents a comparison between standard robustness and top-$k$ robustness, including top-$k$ relaxed robustness and top-$k$ affinity robustness. For each robustness metric, the left/middle/right bars represent the percentage of \unsafe{}, \safe{}, and \timeout{} cases, respectively.
}
\label{fig:combin}
\vspace{-2mm}
\end{figure}

\noindent {\it Relaxed robustness:} As defined in Eq ~\ref{eq:rr} of Section~\ref{sec:examples}, we need a user defined confidence level $\thresh$ to analyse this property.  
We took five different confidence thresholds $\thresh \in \{0, 60, 80, 90, 95\}$, where $\thresh = 0$ corresponds to standard robustness. For a given threshold value $\thresh$, a result of \textsc{safe} indicates either no misclassification or a misclassification with confidence below $\thresh\%$, whereas a result of \textsc{unsafe} indicates a counterexample with confidence above $\thresh\%$.

The results in Figure~\ref{plot:relaxed1} show that as we increase the confidence level, the safe (verified) cases increase and unsafe (counterexamples) decrease. This is because increasing confidence means we report counterexamples only if it has confidence greater than the threshold $\thresh$, which means less numbers of counterexamples. Also, \timeout{} cases are decreasing, as increasing confidence reduces search space, and most of the cases are verified by the abstraction-based module like CROWN~\cite{xu2021fast} and MILP based bound tightening, within \alphabeta.

\noindent{\it Strong robustness:}
We took the threshold value $\thresh_1$ as $57.5$ for the seed image in Eq~\ref{eq:sr}, because it is the average value of all the confidences on the seed images of the properties. We took three different values of the confidences thresholds $\thresh_2=30, 35, 40$. Intuitively we want our model to be robust against the misclassification, at the same time confidence should not go below $\thresh_2$. As shown in Figure~\ref{plot:strong1}, the counterexamples increase as we increase the confidence $\thresh_2$, because higher the threshold means higher chances of confidence to fall below the threshold, with almost no increase in time taken. 

\noindent{\it Smoothness:}  
We considered varying threshold values of $10$, $25$, and $40$ for smoothness. Intuitively, a threshold of $10$ means that if, within an $\epsilon$ perturbation, the confidence of the seed image's class fluctuates within $\pm 10$ of its original confidence, the smoothness property holds (\safe{}); otherwise, it does not hold (\unsafe{}). Figure~\ref{plot:smooth1} illustrates how smoothness changes as the threshold varies. Increasing the smoothness threshold allows for greater variations in the output, resulting in more \safe{} cases and fewer \unsafe{} cases. However, the number of timeout cases slightly increases. A possible reason is that a lower threshold implies even slight changes in output confidence can lead to \unsafe{} cases. 

\noindent{\it Top-k, top-k-relaxed, and top-k-affinity robustness:} In both variations of top-k robustness, no threshold is involved, so the number of benchmarks remains the same as in standard robustness. We observed in previous experiments that for the GTSRB dataset properties, the confidence values for all seed images are $100\%$, meaning the confidence for all other classes is $0\%$. As a result, top-k properties cannot be applied to these benchmarks. Figure~\ref{plot:bargraph} compares standard robustness with top-k, top-k relaxed, and top-k affinity robustness. The number of \unsafe{} cases decreases from standard robustness to the other three robustness definitions. 

Thus, from the above results, we conclude that our layer-based approach, integrated with the state-of-the-art verifier is able to verify all the richer properties that we defined, on large benchmarks from \vnncomp{}, and varying the thresholds results in expected change in number of verified counterexamples.

\subsection{Ablation study datasets-wise} 
\label{app:exp:rq1}
Next, we show a detailed analysis of each definition with respect to different datatsets. 

\subsubsection{Relaxed robustness}
\begin{figure}[t]    
  \centering
  \begin{minipage}{0.45\textwidth}
      \begin{center}
      \scalebox{0.5}{
      \begin{tikzpicture}
    \begin{axis}[
        xlabel= {MNIST}, 
        ylabel={Percentage},
        ylabel style={at={(axis description cs:0.07,0.5)}, anchor=south},
        width=11cm,
        height=8cm,
        xmin=0, xmax=100,
        ymin=0, ymax=100,
        xtick={0, 60, 80, 90, 95},
        ytick={0,20,40,60,80,100},
        legend pos=north west,
        legend entries={\textsc{unsafe}, \textsc{safe}, \textsc{timeout}},
        ymajorgrids=true,
        grid style=dashed,
    ]

    \addplot[
        color=cyan,
        mark=*,
    ]
    coordinates {
        (0,20.0)(60,16.67)(80,0.0)(90,0.0)(95,0.0)
    };

    \addplot[
        color=green,
        mark=*,
        dashed,
    ]
    coordinates {
        (0,70.0)(60,83.33)(80,100)(90,100)(95,100)
    };

    \addplot[
        color=black,
        mark=*,
        dashdotted,
    ]
    coordinates {
        (0,10)(60,0)(80,0)(90,0)(95,0)
    };

    \end{axis}

\end{tikzpicture}
      }
      \end{center}
  \end{minipage}
  \begin{minipage}{0.45\textwidth}
      \begin{center}
      \scalebox{0.5}{
      \begin{tikzpicture}
    \begin{axis}[
        xlabel={CIFAR-10},
        ylabel={Percentage},
        ylabel style={at={(axis description cs:0.07,0.5)}, anchor=south},
        width=11cm,
        height=8cm,
        xmin=0, xmax=100,
        ymin=0, ymax=100,
        xtick={0, 60, 80, 90, 95},
        ytick={0,20,40,60,80,100},
        legend pos=north west,
        legend entries={\textsc{unsafe}, \textsc{safe}, \textsc{timeout}},
        ymajorgrids=true,
        grid style=dashed,
    ]

    \addplot[
        color=cyan,
        mark=*,
    ]
    coordinates {
        (0,14.43)(60,13.11)(80,5.9)(90,2.95)(95,1.97)
    };

    \addplot[
        color=green,
        mark=*,
        dashed,
    ]
    coordinates {
        (0,72.13)(60,75.08)(80,90.82)(90,95.74)(95,96.72)
    };

    \addplot[
        color=black,
        mark=*,
        dashdotted,
    ]
    coordinates {
        (0,13.44)(60,11.80)(80,3.28)(90,1.31)(95,1.31)
    };

    \end{axis}

\end{tikzpicture}
      }
      \end{center}     
  \end{minipage}

  \begin{minipage}{0.45\textwidth}
    \begin{center}
    \scalebox{0.5}{
    \begin{tikzpicture}
    \begin{axis}[
        xlabel={GTSRB},
        ylabel={Percentage},
        ylabel style={at={(axis description cs:0.07,0.5)}, anchor=south},
        width=11cm,
        height=8cm,
        xmin=0, xmax=100,
        ymin=0, ymax=100,
        xtick={0, 60, 80, 90, 95, 98},
        ytick={0,20,40,60,80,100},
        legend pos=north west,
        legend entries={\textsc{unsafe}, \textsc{safe}, \textsc{timeout}},
        ymajorgrids=true,
        grid style=dashed,
    ]

    \addplot[
        color=cyan,
        mark=*,
    ]
    coordinates {
        (0,93.33)(60,93.33)(80,93.33)(90,93.33)(95,93.33)(98,93.33)
    };

    \addplot[
        color=green,
        mark=*,
        dashed,
    ]
    coordinates {
        (0,0)(60,0)(80,0)(90,0)(95,0)(98,0)
    };

    \addplot[
        color=black,
        mark=*,
        dashdotted,
    ]
    coordinates {
        (0,6.67)(60,6.67)(80,6.67)(90,6.67)(95,6.67)(98,6.67)
    };

    \end{axis}

\end{tikzpicture}
    }
    \end{center}
  \end{minipage}
  \begin{minipage}{0.45\textwidth}
    \begin{center}
    \scalebox{0.5}{
    \begin{tikzpicture}
    \begin{axis}[
        xlabel={IMAGENET-1k},
        ylabel={Percentage},
        ylabel style={at={(axis description cs:0.07,0.5)}, anchor=south},
        width=11cm,
        height=8cm,
        xmin=0, xmax=100,
        ymin=0, ymax=100,
        xtick={0, 60, 80, 90, 95},
        ytick={0,20,40,60,80,100},
        legend pos=north west,
        legend entries={\textsc{unsafe}, \textsc{safe}, \textsc{timeout}},
        ymajorgrids=true,
        grid style=dashed,
    ]

    \addplot[
        color=cyan,
        mark=*,
    ]
    coordinates {
        (0,5.56)(60,0)(80,0)(90,0)(95,0)
    };

    \addplot[
        color=green,
        mark=*,
        dashed,
    ]
    coordinates {
        (0,33.33)(60,33.33)(80,33.33)(90,38.89)(95,38.89)
    };

    \addplot[
        color=black,
        mark=*,
        dashdotted,
    ]
    coordinates {
        (0,61.11)(60,66.67)(80,66.67)(90,61.11)(95,61.11)
    };

    \end{axis}

\end{tikzpicture}
    }
    \end{center}
  \end{minipage}
  \caption{
    Analysis of relaxed robustness with respect to each dataset separately. 
  }
  \label{plot:relaxed}
\end{figure}
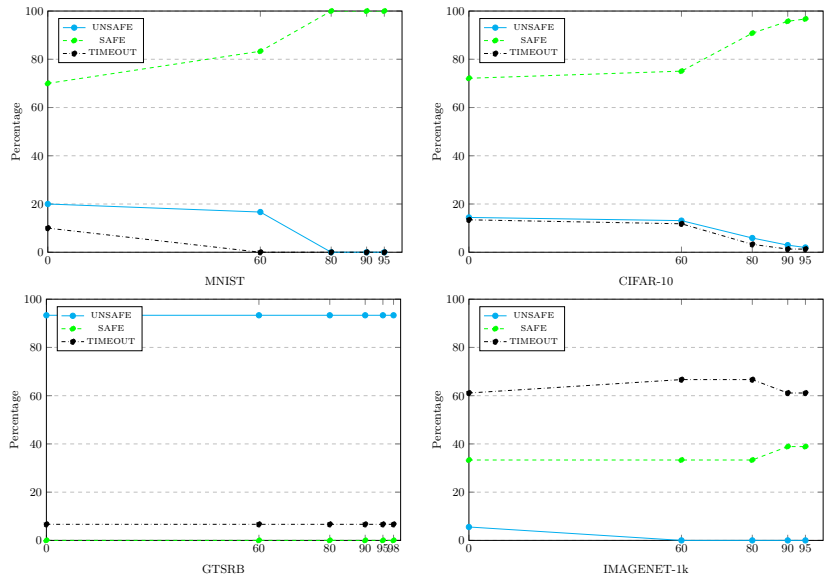

Figure~\ref{plot:relaxed} presents the analysis across different datasets. For the MNIST benchmarks, the number of \safe{} cases increases while the number of \unsafe{} cases decreases as we raise the confidence threshold. This is intuitive, as a higher confidence threshold makes the definition more relaxed, leading to more \safe{} cases. Notably, we observe zero \timeout{} cases beyond a confidence threshold of 80\%. Since these are small networks, \alphabeta{} employs the Gurobi~\cite{gurobioptimizer} solver for verification. As the confidence threshold increases, the search space tightens, allowing all benchmarks to be verified either by CROWN~\cite{xu2021fast} or the MILP bound-tightening technique. A similar trend is observed for the CIFAR-10 benchmarks—higher confidence thresholds make the definitions more relaxed, resulting in more \safe{} cases, which can be efficiently verified by incomplete methods like CROWN.

\begin{figure}
  \centering
  \includegraphics[scale=0.6]{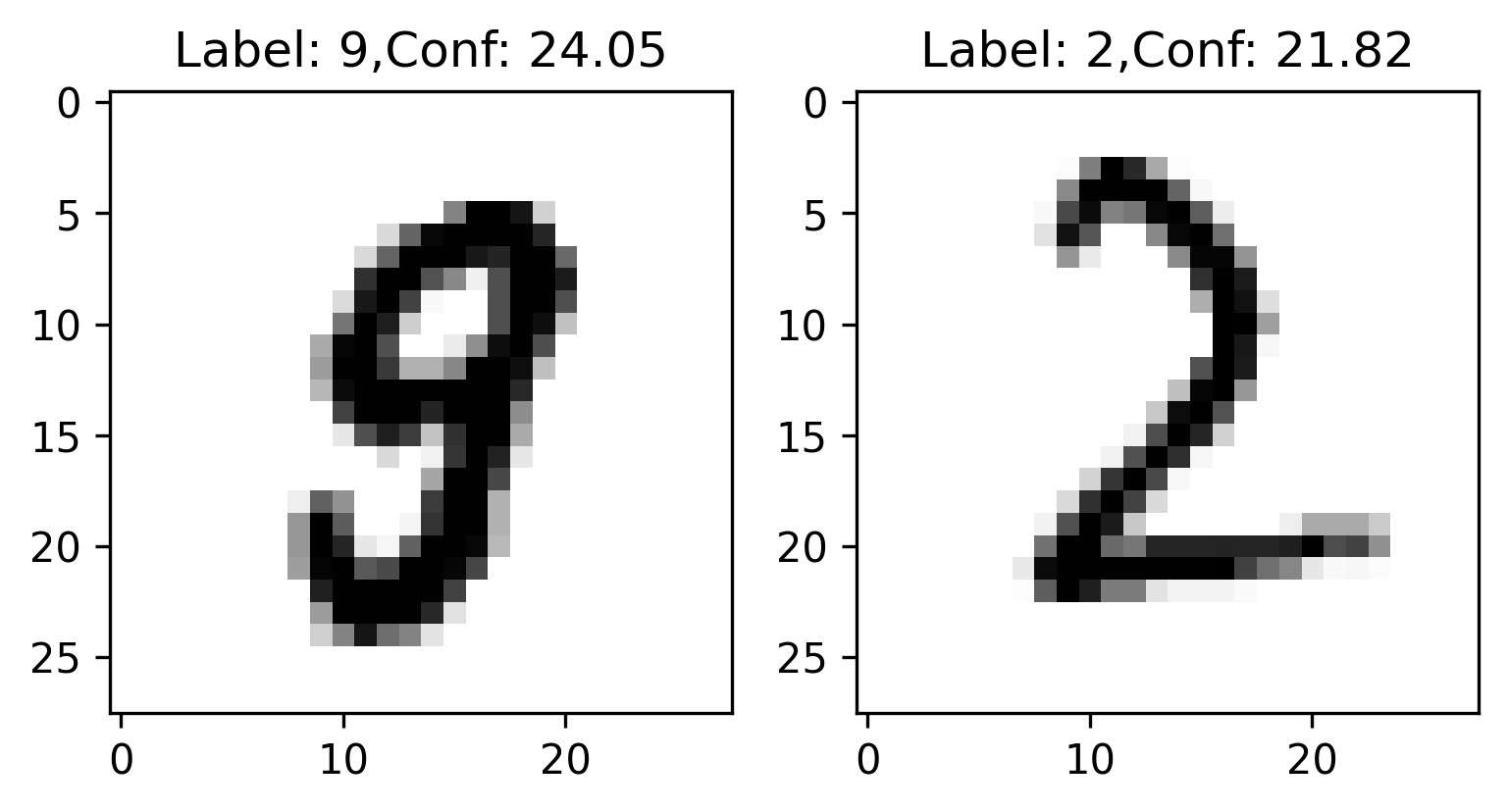}
  \caption{The above images were verified using relaxed robustness with a confidence threshold of $80\%$.}
  \label{fig:relaxed_verified_mnist}
\end{figure}

\begin{figure}
  \centering
  \includegraphics[scale=0.3]{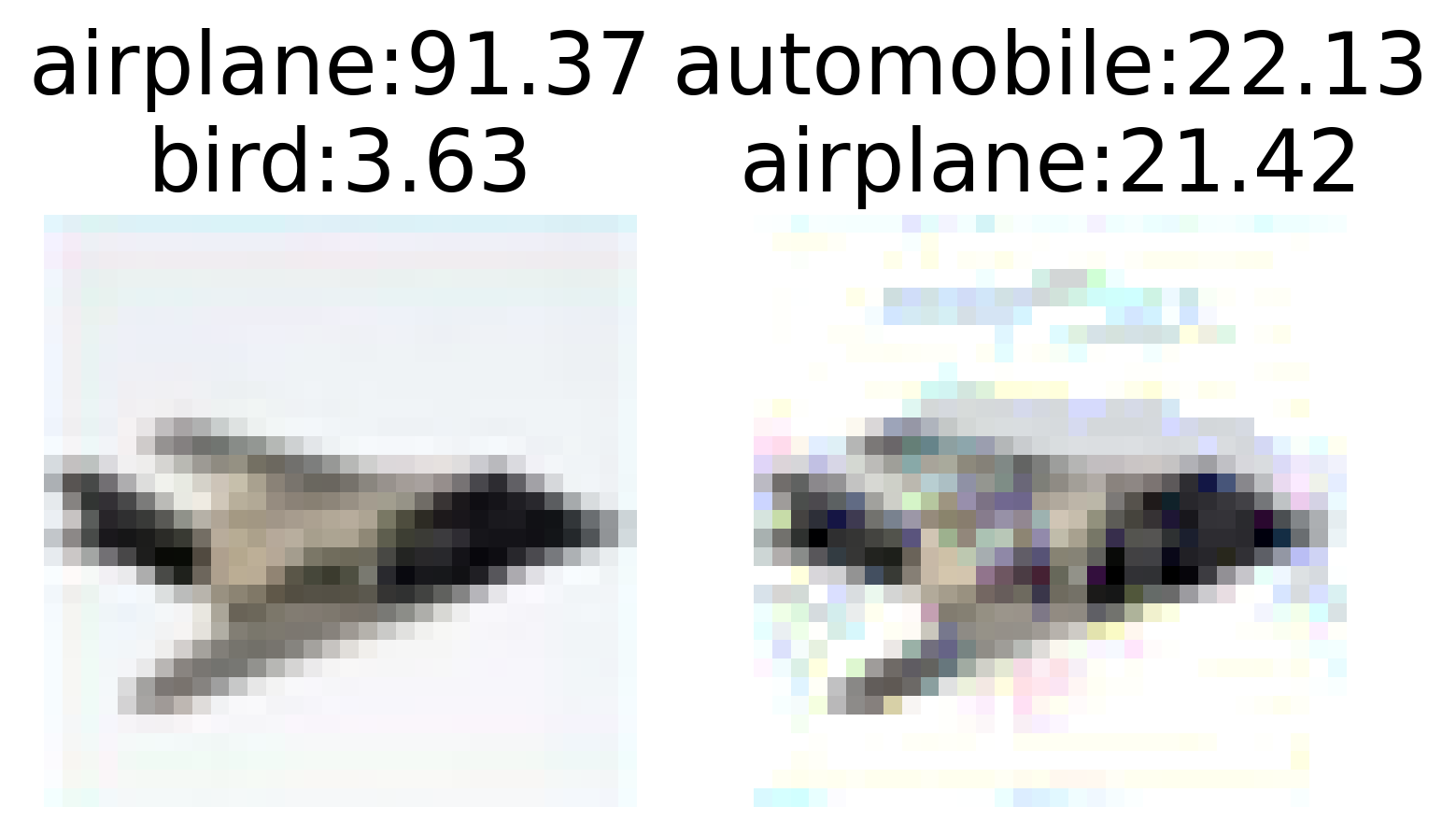}
  \caption{
    The image on the left was correctly classified as label \textsc{airplan} by the network \texttt{convBigRELU-PGD.onnx} with a confidence of $91.37\%$, but it was misclassified as \texttt{automobile} with a confidence of $22.13\%$ within an $\epsilon$ perturbation of $0.007$ under the standard robustness property. The same image was then verified by relaxed robustness with a confidence threshold of $90\%$.
  }
  \label{fig:relaxed_verified_cifar10}
\end{figure}

\begin{figure}[t]
  \centering
      \includegraphics[scale=0.3]{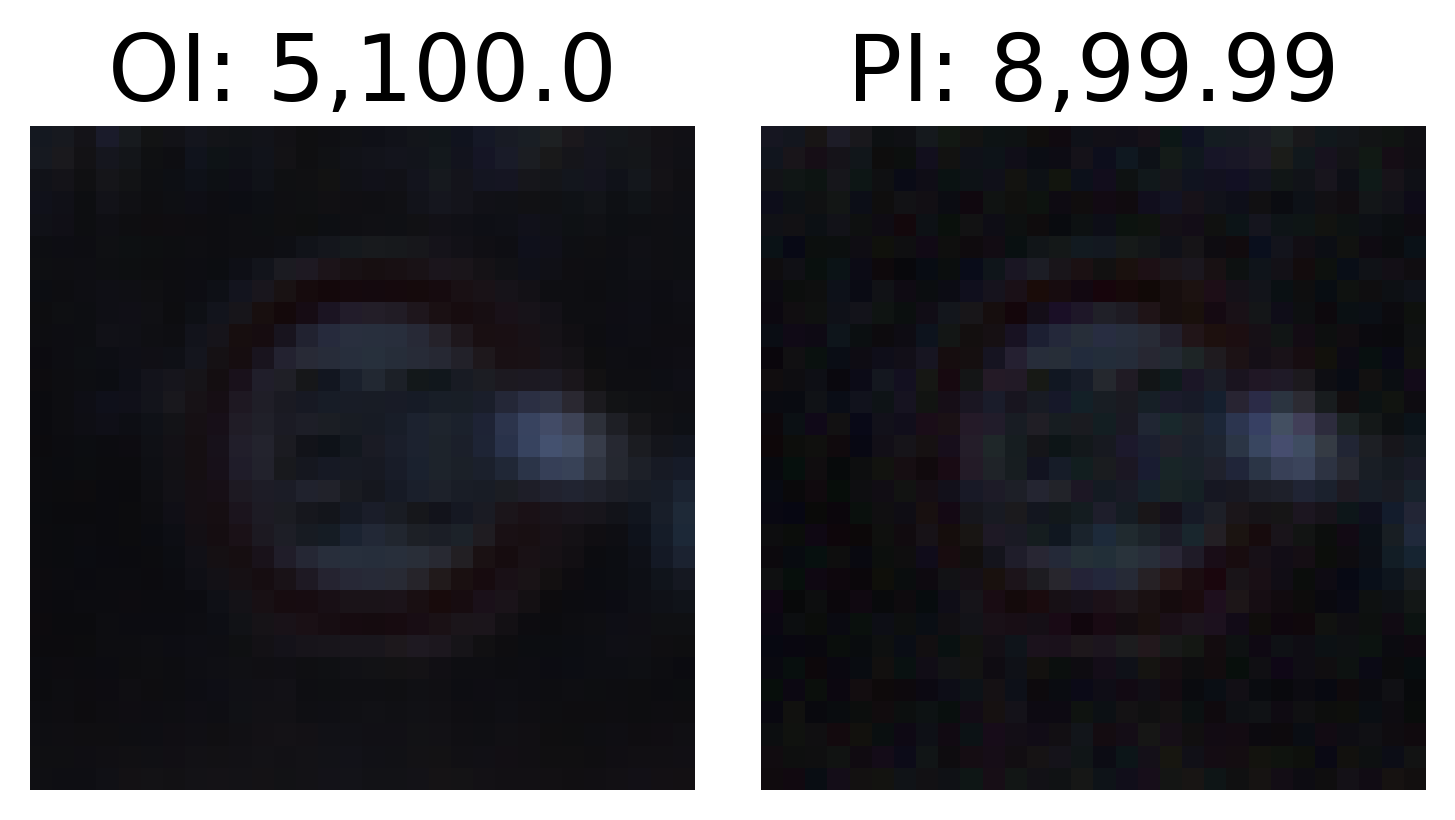}
      \includegraphics[scale=0.3]{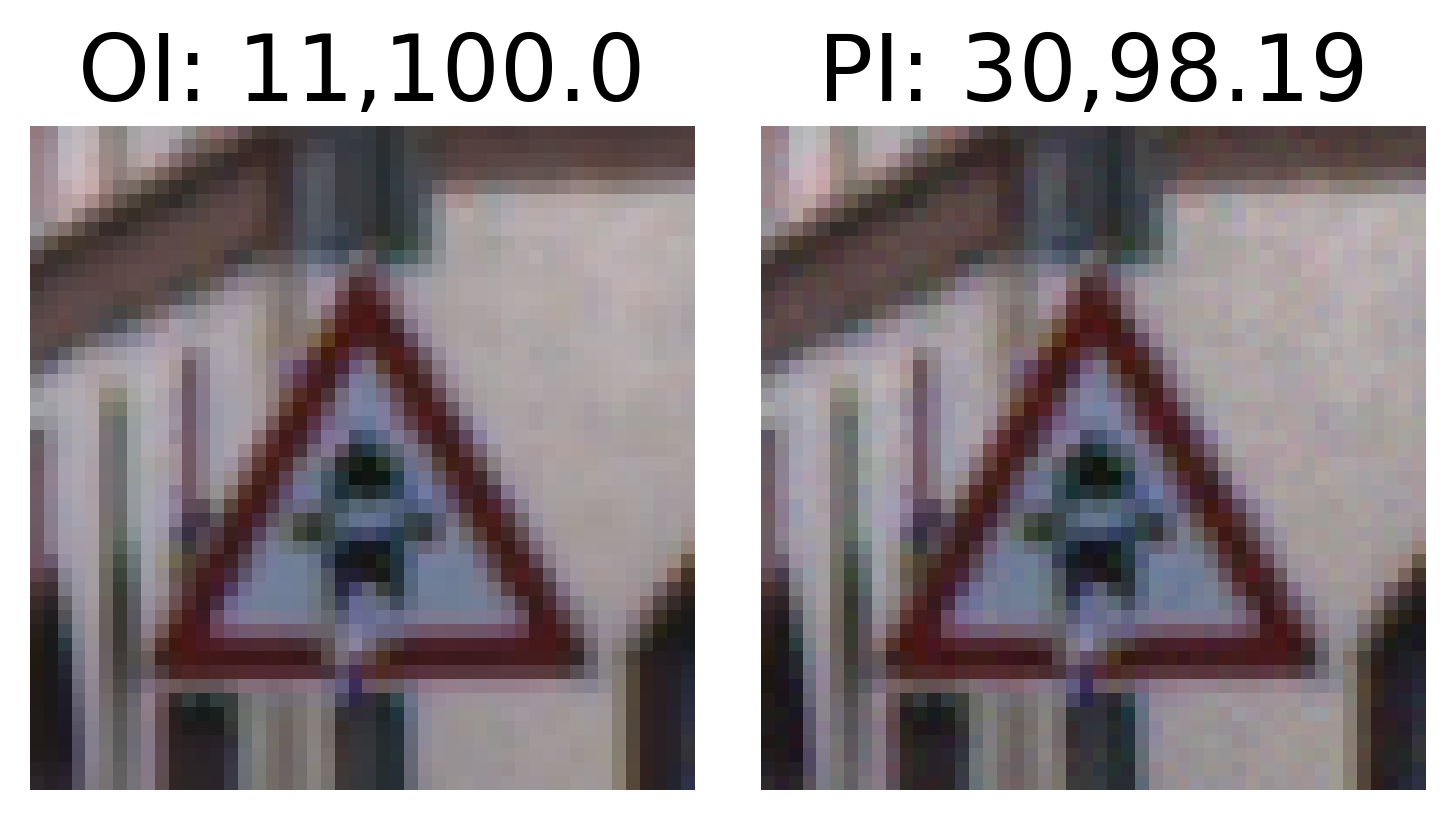}
      \caption{
        The left image is taken from the German Traffic Sign Recognition Benchmark (GTSRB) and belongs to the class \texttt{Speed limit (80 km/h)}. It is correctly classified by \texttt{net-1} (Table~\ref{tab:net_details}) with $100\%$ confidence. However, under an $\epsilon$-perturbation of $5/255$, it is misclassified as \texttt{Speed limit (120 km/h)} with a high confidence of $99.99\%$, highlighting a potential vulnerability in the network.  
        The right image belongs to the class \texttt{Right-of-way at the next intersection} and is classified with $100\%$ confidence. With the same $\epsilon$-perturbation, it is misclassified as \texttt{Beware of ice/snow} with a confidence of $98.19\%$.  
  }
  \label{fig:cex_relaxed}
\end{figure}

\begin{figure}
  \centering
  \includegraphics[scale=0.5]{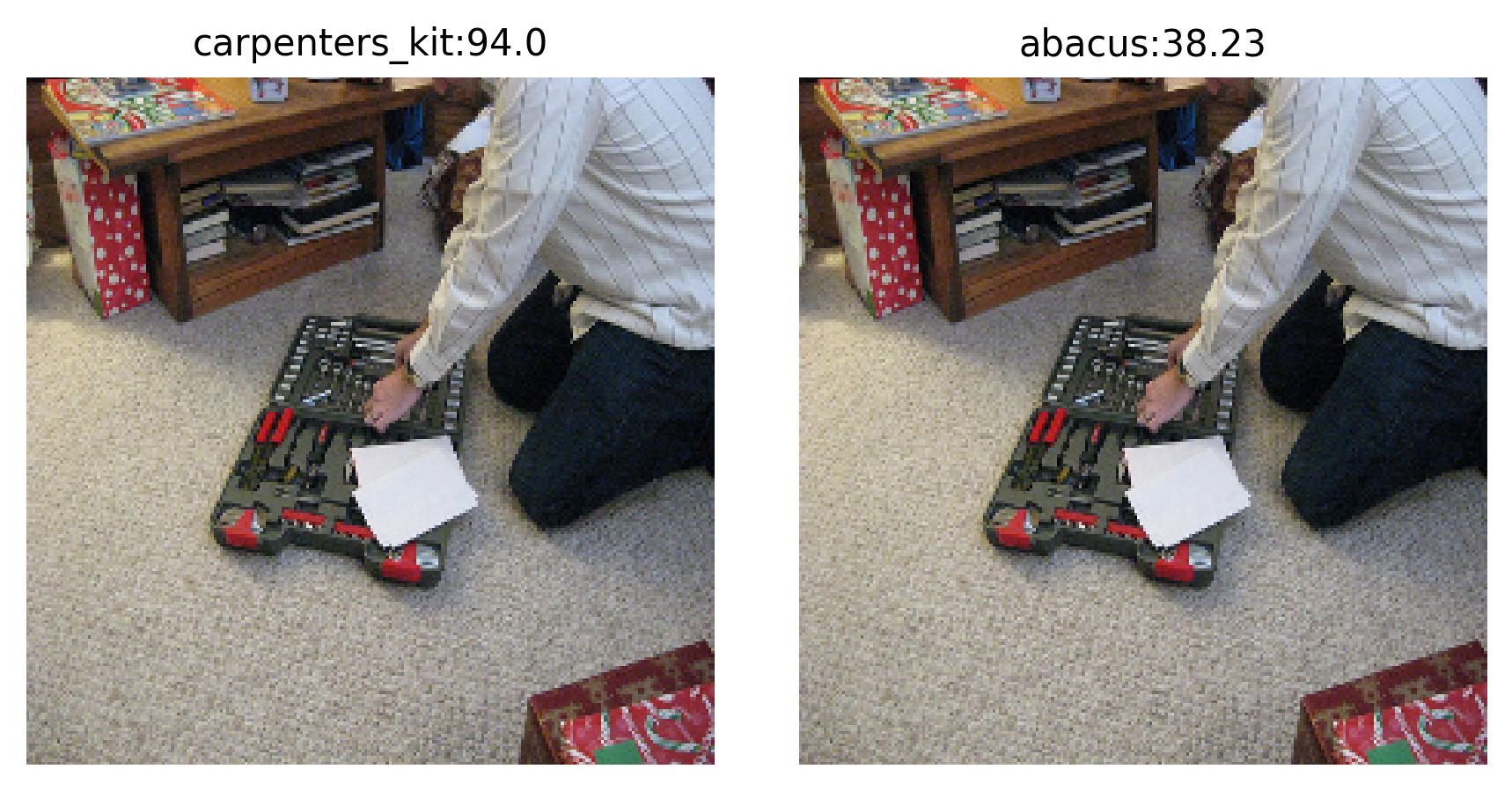}
  \caption{
    The image on the left is correctly classified as \textsc{carpenters kit} with $94.0\%$ confidence by the VGGNET-16 network. However, within an epsilon perturbation of $1e-5$, it is misclassified as \textsc{abacus} with $38.23\%$ confidence. This image was verified under relaxed robustness with a $95\%$ confidence threshold.
  }
  \label{ex:imagenet}
\end{figure}

We observed surprising behavior with the GTSRB dataset, as shown in Figure~\ref{plot:relaxed}. The number of \safe{}, \unsafe{}, and \timeout{} cases remained constant, despite increasing the confidence threshold. There were a total of 45 benchmarks available for this category, resulting in 45 benchmarks for each confidence threshold. For each threshold, we found 42 \unsafe{}, 3 \timeout{}, and 0 \safe{} cases. This behavior raised curiosity, prompting us to experiment with a confidence threshold of 98\%, but the same pattern persisted. Additionally, we noticed that the confidence for all seed images was 100\%. This behavior raises serious concerns about the vulnerability of these networks. If both the seed image and counterexample have near-100\% confidence, it is a red flag for these networks. Confidently making incorrect decisions is more dangerous than doing so with lower confidence, as a user might intervene to validate decisions made with lower confidence.

In the IMAGENET-1K benchmarks, there are a total of 18 benchmarks, implying 18 benchmarks for each confidence threshold. We observed on a GPU machine that many of the benchmarks ran out of memory with 32GB of GPU RAM. As a result, we had to run them on a CPU machine: Intel(R) Xeon(R) Gold 6314U CPU @ 2.30GHz with 64GB of RAM. For the standard robustness property with a confidence of 0\%, we observed 1 \unsafe{}, 6 \safe{}, and 11 \timeout{} cases. At confidence levels of 60\% and 80\%, the 1 \unsafe{} case was converted to a \timeout{}. At 90\% and 95\% confidence, it was converted to a \safe{} case. The 11 \timeout{} cases remained as \timeout{} for all confidence levels. 

Figures~\ref{fig:relaxed_verified_mnist}, \ref{fig:relaxed_verified_cifar10}, and \ref{fig:cex_relaxed}, and \ref{ex:imagenet} provide more insightful examples of the relaxed robustness definition.

\subsubsection{Strong robustness} 

\begin{figure}[t]    
  \centering
  \begin{minipage}{0.30\textwidth}
      \begin{center}
      \scalebox{0.35}{
      \begin{tikzpicture}
    \begin{axis}[
        xlabel= {MNIST}, 
        ylabel={Percentage},
        width=11cm,
        height=8cm,
        xmin=0, xmax=22,
        ymin=0, ymax=100,
        xtick={0, 15, 17, 20},
        ytick={0,20,40,60,80,100},
        legend pos=north west,
        legend entries={\textsc{unsafe}, \textsc{safe}, \text{trivial safe}, \textsc{timeout}},
        ymajorgrids=true,
        grid style=dashed,
    ]

    \addplot[
        color=cyan,
        mark=*,
    ]
    coordinates {
        (0,20.0)(15,84.44)(17,84.44)(20,84.44)
    };

    \addplot[
        color=green,
        mark=*,
        dashed,
    ]
    coordinates {
        (0,70.0)(15,0)(17,0)(20,0)
    };

    \addplot[
        color=gray,
        mark=*,
        dotted,
    ]
    coordinates {
        (0,0)(15,14)(17,14)(20,14)
    };

    \addplot[
        color=black,
        mark=*,
        dashdotted,
    ]
    coordinates {
        (0,10)(15,0)(17,0)(20,0)
    };

    \end{axis}

\end{tikzpicture}
      }
      \end{center}
  \end{minipage}
  \begin{minipage}{0.30\textwidth}
      \begin{center}
      \scalebox{0.35}{
      \begin{tikzpicture}
    \begin{axis}[
        xlabel={CIFAR-10},
        ylabel={Percentage},
        width=11cm,
        height=8cm,
        xmin=0, xmax=42,
        ymin=0, ymax=100,
        xtick={0, 30, 35, 40},
        ytick={0,20,40,60,80,100},
        legend pos=north west,
        legend entries={\textsc{unsafe}, \textsc{safe}, \text{trivial safe}, \textsc{timeout}},
        ymajorgrids=true,
        grid style=dashed,
    ]

    \addplot[
        color=cyan,
        mark=*,
    ]
    coordinates {
        (0,14.43)(30,15.40)(35,19.67)(40,22.30)
    };

    \addplot[
        color=green,
        mark=*,
        dashed,
    ]
    coordinates {
        (0,72.13)(30,25.57)(35,22.95)(40,20.33)
    };

    \addplot[
        color=gray,
        mark=*,
        dotted,
    ]
    coordinates {
        (0,0)(30,41.31)(35,41.21)(40,41.31)
    };

    \addplot[
        color=black,
        mark=*,
        dashdotted,
    ]
    coordinates {
        (0,13.44)(30,17.70)(35,16.07)(40,16.07)
    };

    \end{axis}

\end{tikzpicture}
      }
      \end{center}     
  \end{minipage}
  \begin{minipage}{0.30\textwidth}
    \begin{center}
    \scalebox{0.35}{
    \begin{tikzpicture}
    \begin{axis}[
        xlabel={GTSRB},
        ylabel={Percentage},
        width=11cm,
        height=8cm,
        xmin=0, xmax=42,
        ymin=0, ymax=100,
        xtick={0, 30, 35, 40},
        ytick={0,20,40,60,80,100},
        legend pos=north west,
        legend entries={\textsc{unsafe}, \textsc{safe}, \text{trivial safe}, \textsc{timeout}},
        ymajorgrids=true,
        grid style=dashed,
    ]

    \addplot[
        color=cyan,
        mark=*,
    ]
    coordinates {
        (0,93.33)(30,93.33)(35,93.33)(40,95.56)
    };

    \addplot[
        color=green,
        mark=*,
        dashed,
    ]
    coordinates {
        (0,0)(30,0)(35,0)(40,0)
    };

    \addplot[
        color=gray,
        mark=*,
        dotted,
    ]
    coordinates {
        (0,0)(30,0)(35,0)(40,0)
    };

    \addplot[
        color=black,
        mark=*,
        dashdotted,
    ]
    coordinates {
        (0,6.67)(30,6.67)(35,6.67)(40,4.44)
    };

    \end{axis}

\end{tikzpicture}
    }
    \end{center}
  \end{minipage}
  \caption{
    Analysis of strong robustness with respect to each dataset separately. 
  }
  \label{plot:strong}
\end{figure}
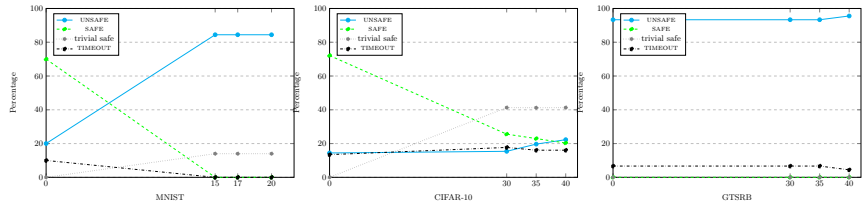

Figure~\ref{plot:strong} shows the strong robustness behavior with varying threshold levels. In addition to \safe{}, \unsafe{}, and \timeout{}, we also introduced a new attribute to the graphs, namely \textsc{trivial safe}. The \textsc{trivial safe} cases occur when the premise of the strong robustness condition in Eq~\ref{eq:sr} becomes false, because the confidence of the seed image is below the threshold. As mentioned in the experimental section of the main paper, we set a 57\% confidence threshold for the seed images, as it represents the average confidence across all seed images and datasets. We observed that all the MNIST benchmarks became trivially safe, as the confidence of all seed images was below 57\%. Therefore, we conducted a separate analysis for the MNIST dataset with varying threshold values.

We took $22\%$ as the threshold on seed images for mnist benchmarks, and threshold values $15$, $17$, and $20$ on the right side thresholds of the strong robustness equation. We found that after confidence threshold $15\%$ the \safe{} and \timeout{} cases become $0$, only \unsafe{} and \textsc{trivial safe} cases remained. This implies these networks are not strong robust with respect to the given threshold values.  

For the CIFAR-10 dataset, we used the same threshold values mentioned in the experimental section of the main paper. As we increased the confidence threshold on the right-hand side of the implication in the strong robustness equation, the number of \unsafe{} cases increased, while the number of \safe{} cases decreased. This trend is intuitive, as a higher confidence requirement increases the likelihood of falsifying the condition. The number of \timeout{} cases remained almost unchanged.

We used the same threshold values as those for the CIFAR-10 dataset for the GTSRB dataset. The number of \textsc{trivial safe} cases is zero since all seed images have a confidence of $100\%$. We observed almost the same behavior as in relaxed robustness, except for one case that converted from \timeout{} to \unsafe{} at the confidence threshold of $40\%$.  

\begin{figure}
  \centering
  \includegraphics[scale=0.3]{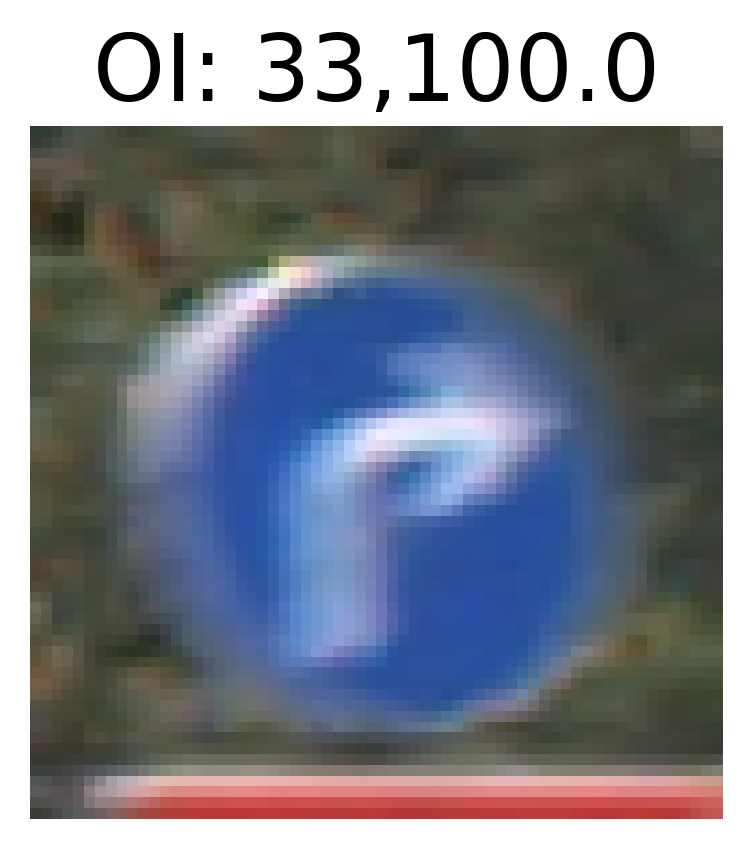}
  \caption{
    This image from the GTSRB benchmark was classified correctly \texttt{Turn right ahead (33)} with $100\%$ confidence by the network $net-3$ in Table~\ref{tab:net_details}. However, this case resulted in a timeout in all three definitions: relaxed robustness, strong robustness, and smoothness.
}
  \label{fig:timeout_gtsrb}
\end{figure}

\subsubsection{Smoothness}  
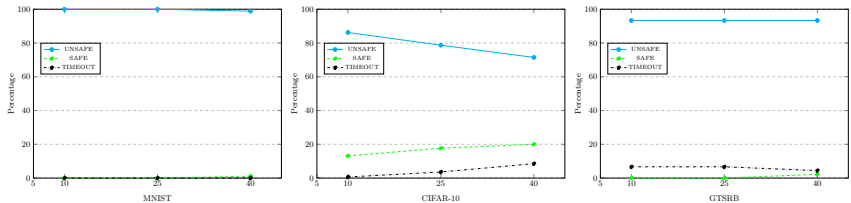
\begin{figure}[t]    
  \centering
  \begin{minipage}{0.30\textwidth}
      \begin{center}
      \scalebox{0.35}{
      \begin{tikzpicture}
    \begin{axis}[
        xlabel= {MNIST}, 
        ylabel={Percentage},
        ylabel style={at={(axis description cs:0.07,0.5)}, anchor=south},
        width=11cm,
        height=8cm,
        xmin=5, xmax=45,
        ymin=0, ymax=100,
        xtick={5, 10, 25, 40},
        ytick={0,20,40,60,80,100},
        legend style={at={(0.27,0.8)}, anchor=north east},  
        legend entries={\textsc{unsafe}, \textsc{safe}, \textsc{timeout}},
        ymajorgrids=true,
        grid style=dashed,
    ]

    \addplot[
        color=cyan,
        mark=*,
    ]
    coordinates {
        (10,100)(25,100)(40,98.88)
    };

    \addplot[
        color=green,
        mark=*,
        dashed,
    ]
    coordinates {
        (10,0)(25,0)(40,1.12)
    };

    \addplot[
        color=black,
        mark=*,
        dashdotted,
    ]
    coordinates {
        (10,0)(25,0)(40,0)
    };

    \end{axis}

\end{tikzpicture}
      }
      \end{center}
  \end{minipage}
  \begin{minipage}{0.30\textwidth}
      \begin{center}
      \scalebox{0.35}{
      \begin{tikzpicture}
    \begin{axis}[
        xlabel= {CIFAR-10}, 
        ylabel={Percentage},
        ylabel style={at={(axis description cs:0.07,0.5)}, anchor=south},
        width=11cm,
        height=8cm,
        xmin=5, xmax=45,
        ymin=0, ymax=100,
        xtick={5, 10, 25, 40},
        ytick={0,20,40,60,80,100},
        legend style={at={(0.27,0.8)}, anchor=north east},  
        legend entries={\textsc{unsafe}, \textsc{safe}, \textsc{timeout}},
        ymajorgrids=true,
        grid style=dashed,
    ]

    \addplot[
        color=cyan,
        mark=*,
    ]
    coordinates {
        (10,86.23)(25,78.69)(40,71.47)
    };

    \addplot[
        color=green,
        mark=*,
        dashed,
    ]
    coordinates {
        (10,13.11)(25,17.70)(40,20)
    };

    \addplot[
        color=black,
        mark=*,
        dashdotted,
    ]
    coordinates {
        (10,0.65)(25,3.60)(40,8.52)
    };

    \end{axis}

\end{tikzpicture}
      }
      \end{center}     
  \end{minipage}
  \begin{minipage}{0.30\textwidth}
    \begin{center}
    \scalebox{0.35}{
    \begin{tikzpicture}
    \begin{axis}[
        xlabel= {GTSRB}, 
        ylabel={Percentage},
        ylabel style={at={(axis description cs:0.07,0.5)}, anchor=south},
        width=11cm,
        height=8cm,
        xmin=5, xmax=45,
        ymin=0, ymax=100,
        xtick={5, 10, 25, 40},
        ytick={0,20,40,60,80,100},
        legend style={at={(0.27,0.8)}, anchor=north east},  
        legend entries={\textsc{unsafe}, \textsc{safe}, \textsc{timeout}},
        ymajorgrids=true,
        grid style=dashed,
    ]

    \addplot[
        color=cyan,
        mark=*,
    ]
    coordinates {
        (10,93.33)(25,93.33)(40,93.33)
    };

    \addplot[
        color=green,
        mark=*,
        dashed,
    ]
    coordinates {
        (10,0)(25,0)(40,2.22)
    };

    \addplot[
        color=black,
        mark=*,
        dashdotted,
    ]
    coordinates {
        (10,6.67)(25,6.67)(40,4.44)
    };

    \end{axis}

\end{tikzpicture}
    }
    \end{center}
  \end{minipage}
  \caption{
    Analysis of smoothness with respect to each dataset separately. 
  }
  \label{plot:smooth}
\end{figure}

Figure~\ref{plot:smooth} illustrates the behavior of the smoothness property across all three datasets. We observed that for the MNIST and GTSRB datasets, almost 99\% of the benchmarks did not satisfy the smoothness property, indicating highly unsmooth behavior in these networks. The Figure~\ref{fig:timeout_gtsrb} shows the timeout case. The CIFAR-10 benchmarks performed reasonably well, with the number of \safe{} cases increasing as the smoothness threshold increased. This trend is intuitive since a higher smoothness threshold allows for greater variations in the output. Notably, we observed a slight increase in \timeout{} cases, which may be due to the higher complexity of this property compared to relaxed and strong robustness. Unlike the other properties, smoothness requires verifying both upper and lower bounds of the output confidence.

\subsubsection{Top-k Robustness}
\label{app:exp:topk}

In both variations of top-k robustness, no threshold is involved, so the number of benchmarks remains the same as in standard robustness. We observed in previous experiments that for the GTSRB dataset properties, the confidence values for all seed images are $100\%$, meaning the confidence for all other classes is $0\%$. As a result, top-k properties cannot be applied to these benchmarks. 

\begin{figure}[t]    
  \centering
  \begin{minipage}{0.45\textwidth}
      \begin{center}
      \scalebox{0.55}{
      \begin{tikzpicture}
    \begin{axis}[
        width=10.2cm, height=7cm,
        ybar, 
        bar width=10pt, 
        ymin=0, ymax=100, 
        symbolic x coords={Standard, top-k, top-k-relaxed, top-k-affinity}, 
        xtick=data,
        nodes near coords, 
        every node near coord/.append style={black}, 
        enlarge x limits=0.2, 
        ylabel={Percentage},
        ylabel style={at={(axis description cs:0.1,0.5)}, anchor=south},
        xlabel={MNIST},
        legend style={at={(0.15,1.0)}, anchor=north}, 
        xticklabel style={rotate=0, anchor=center}, 
        ymajorgrids=true, grid style=dashed, 
        x tick label style={yshift=-5pt}, 
    ]
    
    \addplot[color=yellow, fill=yellow, bar shift=-15pt] coordinates {
        (Standard,20) (top-k,23.40) (top-k-relaxed,10) (top-k-affinity,17.78)
    };
    
    \addplot[color=cyan, fill=cyan, bar shift=0pt] coordinates {
        (Standard,70) (top-k, 68.12) (top-k-relaxed,82.22) (top-k-affinity,78.88)
    };
    
    \addplot[color=green, fill=green, bar shift=15pt] coordinates {
        (Standard,10) (top-k,8.48) (top-k-relaxed,7.78) (top-k-affinity,3.33)
    };
    
    \legend{\unsafe{}, \safe{}, \timeout{}}
    
    \end{axis}
\end{tikzpicture}
      }
      \end{center}
  \end{minipage}
  \begin{minipage}{0.45\textwidth}
    \begin{center}
    \scalebox{0.55}{
    \begin{tikzpicture}
    \begin{axis}[
        width=10.2cm, height=7cm,
        ybar, 
        bar width=10pt, 
        ymin=0, ymax=100, 
        symbolic x coords={Standard, top-k, tok-k-relaxed, top-k-affinity}, 
        xtick=data,
        nodes near coords, 
        every node near coord/.append style={black}, 
        enlarge x limits=0.2, 
        ylabel={Percentage},
        ylabel style={at={(axis description cs:0.1,0.5)}, anchor=south},
        xlabel={CIFAR-10},
        legend style={at={(0.85,1.0)}, anchor=north}, 
        xticklabel style={rotate=0, anchor=center}, 
        ymajorgrids=true, grid style=dashed, 
        x tick label style={yshift=-5pt}, 
    ]
    
    \addplot[color=yellow, fill=yellow, bar shift=-15pt] coordinates {
        (Standard,14.43) (top-k, 13.48) (tok-k-relaxed,9.18) (top-k-affinity,9.51) 
    };
    
    \addplot[color=cyan, fill=cyan, bar shift=0pt] coordinates {
        (Standard,72.13) (top-k, 45.20) (tok-k-relaxed,51.48) (top-k-affinity,53.11)
    };
    
    \addplot[color=green, fill=green, bar shift=15pt] coordinates {
        (Standard,13.44) (top-k, 41.32) (tok-k-relaxed,39.34) (top-k-affinity,37.38) 
    };
    
    \legend{\unsafe{}, \safe{}, \timeout{}}
    
    \end{axis}
\end{tikzpicture}
    }
    \end{center}
  \end{minipage}
  \caption{Top-K: The left figure shows the comparison on MNIST benchmarks, right figure shows comparison on CIFAR-10 benchmarks. This figure presents a comparison between standard robustness and top-$k$ robustness, including top-$k$ relaxed robustness and top-$k$ affinity robustness. For each robustness metric, the left/middle/right bars represent the percentage of \unsafe{}, \safe{}, and \timeout{} cases, respectively.}
  \label{plot:topk}
\end{figure}

Affinity robustness requires a predefined set of classes in which misclassification is allowed (prior knowledge). For MNIST, this affinity set is defined as:  \{\{0,8\}, \{4,9\}, \{1,9,7\},\{2\}, \{3\}, \{5\}, \{6\}\}. Intuitively, this means that $0$ is allowed to be misclassified as $8$ but not as any other class, while $9$ is allowed to be misclassified as $4$, $1$, or $7$ only. Similarly, $2$ is not allowed to be misclassified into any other class.  For CIFAR-10, we define the affinity set as:  
\{\{bird\}, \{airplane, automobile, ship, truck\},\{cat\}, \{deer, dog, horse\}, \{dog\}, \{frog\}\}.  
Here, machines (airplane, automobile, ship, truck) are allowed to be classified among themselves, and the same applies to animals (deer, dog, horse). The affinity set is user-defined and may vary depending on the user or application. We took $k=2$ for both MNIST and CIFAR-10 benchmarks. 

Figure~\ref{plot:topk} compares the results on the MNIST and CIFAR-10 datasets separately. As mentioned in the experiment section of the main paper, all the seed images in the GTSRB dataset have $100\%$ confidence, making top-k analysis inapplicable to those benchmarks.  

For the MNIST benchmarks, the results align with intuition: \safe{} cases increase while \unsafe{} cases decrease as we move from standard robustness to affinity and then to top-k relaxed robustness. This trend is likely due to the relatively small network sizes, making it easier to find solutions.  

On the other hand, for the CIFAR-10 benchmarks, while \safe{} cases are theoretically expected to increase, we observed a rise in \timeout{} cases instead. A potential reason for this could be the increased complexity of the problem, specifically the presence of conjunctions of disjunctions, making verification more computationally challenging.

\end{document}